\documentstyle[eqsecnum,prd,aps,epsf]{revtex}

\begin{document}

\def\beqa{\begin{eqnarray}}
\def\eeqa{\end{eqnarray}}
\def\bI{\hbox{$\,I\!\!\!\!-$}}
\def\e{\epsilon}
\def\O{\Omega}
\def\L{\Lambda}
\def\l{\lambda}
\def\D{\Delta}
\def\d{\delta}
\def\r{\rho}
\def\a{\alpha}
\def\b{\beta}
\def\g{\gamma}
\def\k{\kappa}
\def\s{\sigma}
\def\vp{\varphi}
\def\t{\tilde}
\def\th{\theta}
\def\p{\partial}
\def\ra{\rightarrow}
\def\lsim{\mathrel{\mathpalette\Oversim<}}
\def\gsim{\mathrel{\mathpalette\Oversim>}}
\def\Oversim#1#2{\lower0.5ex\vbox{\baselineskip0pt\lineskip0pt%
            \lineskiplimit0pt\ialign{%
          $\mathsurround0pt #1\hfil##\hfil$\crcr#2\crcr\sim\crcr}}}

\thispagestyle{empty}
{\baselineskip0pt
\leftline{\large\baselineskip16pt\sl\vbox to0pt{\hbox{\it Department of Physics}
               \hbox{\it Kyoto University}\vss}}
\rightline{\baselineskip16pt\rm\vbox to20pt{\hbox{KUNS 1540}
           \hbox{January, 1999}
\vss}}%
}
\vskip0.5cm

\begin{center}
{\large \bf
Irrotational and Incompressible Binary Systems \\
in the First Post-Newtonian Approximation of General Relativity
}
\end{center}

\vskip 0.4cm

\begin{center}
Keisuke Taniguchi
\footnote{Electronic address: taniguci@tap.scphys.kyoto-u.ac.jp,}
\end{center} 

\begin{center}
{\em Department of Physics,~Kyoto University,~Kyoto~606-8502,~Japan}
\end{center}

\begin{abstract}
  The first post-Newtonian (PN) hydrostatic equations for an
  irrotational fluid are solved for an incompressible binary system. The
  equilibrium configuration of the binary system is given by a small
  deformation from the irrotational Darwin-Riemann ellipsoid which is
  the solution at Newtonian order. It is found that the orbital
  separation at the innermost stable circular orbit (ISCO) decreases
  when one increases the compactness parameter $M_{\ast}/c^2 a_{\ast}$,
  in which $M_{\ast}$ and $a_{\ast}$ denote the mass and the radius of a
  star, respectively.  If we compare the 1PN angular velocity of the
  binary system at the ISCO in units of $\sqrt{M_{\ast}/a_{\ast}^3}$
  with that of Newtonian order, the angular velocity at the ISCO is
  almost the same value as that at Newtonian order when one increases
  the compactness parameter. Also, we do not find the instability point
  driven by the deformation at 1PN order, where a new sequence
  bifurcates throughout the equilibrium sequence of the binary system
  until the ISCO.

  We also investigate the validity of an ellipsoidal approximation, in
  which a 1PN solution is obtained assuming an ellipsoidal figure and
  neglecting the deformation. It is found that the ellipsoidal
  approximation gives a fairly accurate result for the total energy,
  total angular momentum and angular velocity. However, if we neglect
  the velocity potential of 1PN order, we tend to overestimate the
  angular velocity at the ISCO regardless of the shape of the star
  (ellipsoidal figure or deformed figure).
\end{abstract}

\section{Introduction}

At the beginning of the next century, we will have powerful instruments
for the detection of gravitational waves, such as LIGO,\cite{LIGO}
VIRGO,\cite{VIRGO} GEO\cite{GEO} and TAMA.\cite{TAMA} One of the most
promising sources of gravitational waves in the sensitive frequency
range of these laser interferometers is coalescing binary neutron stars
(BNSs). If we have accurate theoretical templates of inspiraling phase
of BNSs, we can extract various information regarding them from
gravitational waves, such as their mass and spin.\cite{Thorne} Moreover,
when we construct reliable theoretical templates around the innermost
stable circular orbit (ISCO) of BNSs, we can obtain physical information
about the equation of state of neutron stars, i.e., the relation between
the mass and the radius of a neutron star.\cite{lindblom}

Binary neutron stars evolve due to the radiation reaction of
gravitational waves, so that they cannot reach equilibrium states.
However, the timescale of the orbital decay is much longer than the
orbital period of BNSs until the ISCO. Therefore, we can regard the
state of a binary system as a quasi-equilibrium, even if the orbit
approaches the ISCO.

{}From this point of view, several authors have attempted to obtain
relativistic quasi-equilibrium configurations of BNSs
numerically.\cite{shibata,BCSST} However, they assume that the binary
system is synchronized. Synchronization (or corotation) is not a
realistic assumption for the BNSs velocity field just outside the ISCO.
This is because the viscosity of a neutron star is negligible even near
the ISCO, and the velocity field of BNSs becomes irrotational or nearly
irrotational.\cite{Kochanek,BC}

Using the Newtonian theory, Ury\=u and Eriguchi have recently
investigated an irrotational binary system.\cite{UE1,UE2,UE3} However,
since BNSs are general relativistic objects, the Newtonian treatment is
not sufficient and we need to include general relativistic effects. In
order to investigate general relativistic effects on the orbital motion
of BNSs, equilibrium sequences of BNSs composed of Newtonian stars with
interaction forces of 2PN order have been studied.\cite{TNLWOK} On the
other hand, Wilson, Mathews and Marronetti have computed sequences of
irrotational BNSs in the conformally flat approximation of general
relativity.\cite{WMM} They suggest that the central densities of the
stars increase when the stars decrease their separation and massive
neutron stars collapse to black holes prior to merger. This conclusion
is in contradicting with that of other works.\cite{density,BCSST,Lom}
Therefore, it is necessary to solve the quasi-equilibrium problem in
general relativity for an irrotational binary system.

In order to obtain accurate theoretical templates of gravitational waves
from irrotational BNSs just prior to the ISCO, which is our goal, a
general relativistic hydrostatic problem with compressible equation of
state must be numerically solved. Even if we could obtain such numerical
solutions, it is necessary to compare them with analytic or
semi-analytic ones in order to check the validity of the numerical
calculation. Lombardi, Rasio and Shapiro have semi-analytically studied
irrotational BNSs using the energy variational method in the first
post-Newtonian (PN) approximation of general relativity.\cite{Lom}
However, they do not solve the velocity fields of the binary system at
1PN order. They only give the velocity fields of the incompressible
fluid at Newtonian order which flows along a plane perpendicular to the
rotational axis $x_3$. We know that the existence of the $x_3$ component
of the velocity field is important in the irrotational fluid
problem.\cite{UE1,UE2,UE3,TAS} {}From this fact, we believe the velocity
potential of 1PN order in the irrotational binary problem is
important.

We have already determined the equilibrium sequence of a corotating
binary system.\cite{TS} However, the formalism which we have used in
these papers is not applicable for the irrotational case. A formalism
for obtaining quasi-equilibrium configuration of an irrotational
binary system in general relativity has recently been
constructed.\cite{BGM,Asada,Shibata98,Teukolsky,Gourgoulhon} In this
formalism, we need to solve only two hydrostatic equations as for the
fluid equations. One of them consists of the integrated forms of the
Euler equation and the other is the Poisson equation for the velocity
potential. Thus, the formalism seems to be very tractable for
computing the equilibrium configuration of realistic irrotational bodies. In
order to develop the method for solving the irrotational binary problem
consistently, the present author and collaborators recently investigated
an irrotational and incompressible star of 1PN order,\cite{TAS} using
above mentioned formalism. Our method was originally introduced by
Chandrasekhar.\cite{chandra65,chandra67,chandra71,chandra74} As an
extension of this method, we study an irrotational and incompressible
binary system in the 1PN approximation of general relativity in this
paper. We assume that each star in the binary system is constructed from
an incompressible and homogeneous fluid. This assumption presents a
great advantage for solving equations analytically. Also, the
deformation of the figure of the binary system at 1PN order is given by
the Lagrangian displacement vectors.

This paper is organized as follows. In \S 2, we formulate the method to
solve the irrotational binary problem. In \S 3, the Lagrangian
displacement vectors for description of the deformation of the binary
system are given, and the angular velocity of 1PN order is derived from
the first tensor virial relation. We give the boundary conditions for
determining the velocity potential and the deformed figure of 1PN order
in \S 4. The total energy and total angular momentum of the binary
system are calculated in \S 5 and numerical results are given in \S 6.
Section 7 is devoted to summary and discussion.

Throughout this paper, $c$ denotes the light velocity and we use units
in which $G=1$. Latin indices $i$, $j$, $k$, $\cdots$ take values 1 to
3, and $\d_{ij}$ denotes the Kronecker delta.  We use $I_{ij}$ and
$\bI_{ij}$ as the quadrupole moment and its trace free part,
\beqa
  \bI_{ij} =I_{ij} -{1 \over 3} \d_{ij} \sum_{k=1}^3 I_{kk}
\eeqa
of each star in the binary system.

\section{Formulation}

Equilibrium configurations of binary systems with non-uniform velocity
fields are obtained by solving the Euler, continuity and Poisson
equations consistently. Since we consider a binary system composed of
incompressible stars, all the calculations are carried out analytically,
even in the 1PN case. The procedure is as follows.

\noindent
(1) We construct a Newtonian equilibrium configuration of the binary
system, i.e., the irrotational Darwin-Riemann
ellipsoid,\cite{chandra69,Aizenman,LRS94} as a non-perturbed state. For
simplicity, we consider only the case in which the directions of the
vorticity vectors and the angular velocity vector lie along $x_3$-axis,
and we assume that BNSs are composed of equal mass stars.

\noindent
(2) The 1PN corrections for the velocity potential and gravitational
potentials are obtained from their 1PN Poisson equations.

\noindent
(3) We calculate the deformation from the Newtonian binary system
induced by 1PN gravity using the Lagrangian displacement vectors
introduced by Chandrasekhar.\cite{chandra69} Then, the corrections for
the Newtonian quantities due to the deformation of the binary system are
estimated.

\noindent
(4) From the first tensor virial relation at 1PN order, i.e., the force
balance equation at 1PN order, we calculate the correction for the
angular velocity.\cite{TS}

\noindent
(5) We substitute all the 1PN corrections obtained in (2), (3) and (4)
into the 1PN Euler and continuity equations. Then, the coefficients of the
Lagrangian displacement vectors and the 1PN velocity potential are
determined from these two equations with the boundary conditions on the
stellar surface.

In this section, we calculate the Newtonian and 1PN terms which we need
in the above procedures. We consider the equilibrium sequences of BNSs
of equal masses ($M_1 =M_2 =M$) whose coordinate separation is
$R$.\footnote{The coordinate condition in this paper is the standard PN
  one.\cite{chandra65}} For calculational simplicity, we use four kinds
of coordinate systems which are all rotating frames. The first frame is
$x_i$, in which the center of mass of a star (star 1) is located at the
origin of the coordinate system, and the other one (star 2) is located
at $(x_1,~x_2,~x_3)=(-R,~0,~0)$ (see Fig. 1). The second one is $y_i$,
in which the center of mass of star 2 is located at the origin, and star
1 is located at $(y_1,~y_2,~y_3)=(R,~0,~0)$. The third one is $X_i$, in
which the origin is located at the center of mass of the binary system,
and the centers of masses of two stars are located at
$(X_1,~X_2,~X_3)=(R/2,~0,~0)$ and $(-R/2,~0,~0)$. The fourth one is
$Y_i$. This frame is just the same rotating frame as $X_i$, but we label
it with a different name for convenience. We use $X_i$ for star 1 and
$Y_i$ for star 2.  Then, the relations among the four corotating frames
are
\beqa
  \bigl( X_1,~X_2,~X_3 \bigr) &=& \Bigl( x_1 +{R \over 2},~x_2,~x_3
  \Bigr), \\
  \bigl( Y_1,~Y_2,~Y_3 \bigr) &=& \Bigl( y_1 -{R \over 2},~y_2,~y_3
  \Bigr).
\eeqa
Due to the symmetry, we consider only the equilibrium configuration of
star 1 in the following.

\begin{figure}[h,t]
\epsfxsize 10cm 
\begin{center}
\leavevmode
\epsfbox{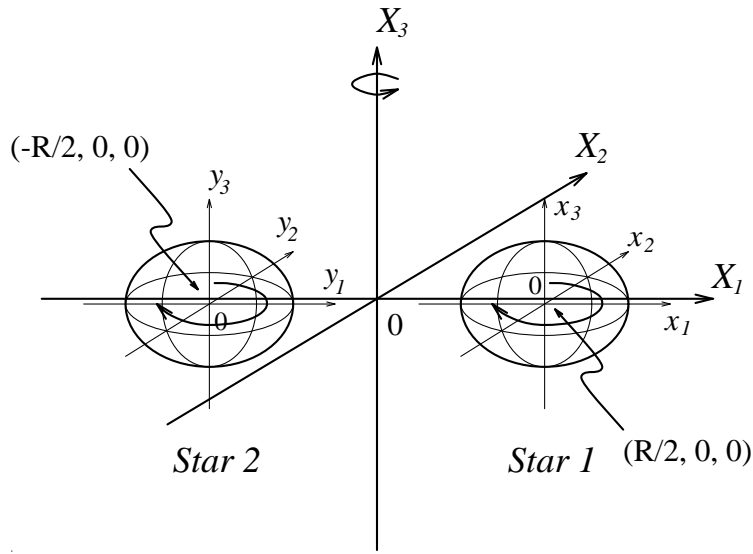}
\end{center}
\caption{Sketch of the Darwin-Riemann ellipsoid. Three coordinate
  systems are shown. The center of mass of star 1 is located at the
  origin of the corotating frame $x_i$, and that of star 2 is located at
  the origin of $y_i$. These two origins of the corotating frames are
  located at $(R/2, 0, 0)$ and $(-R/2, 0, 0)$ in the coordinate system
  $X_i$ in which the origin is located at the center of mass of the
  binary system.
}
\end{figure}

The main purpose of this paper is to calculate the 1PN corrections for
the angular velocity, the velocity potential, the energy, and the
angular momentum in the Darwin-Riemann problem. In the calculation, we
assume that the parameter $a_0/R$, where $a_0$ is a typical radius of a
star, is small and we use it as the expansion parameter. Accordingly,
there exist two different types of expansion parameters in this paper.
One of them is the post-Newtonian parameter and the other is $a_0/R$.

At Newtonian order, the angular velocity ($\O_{\rm DR}$)
becomes\cite{LRS93} (see \S \ref{secangvelo} for derivation)
\beqa
  \O_{\rm DR}^2 ={2M \over R^3} +{18\bI_{11} \over R^5}
  +O(R^{-7}). \label{Nomega}
\eeqa
Thus, in the 1PN approximation, we can expect that the following types of 
quantities will be the main terms at 1PN order:
\beqa
  \sim {M \over R^3} \times {M \over a_0 c^2},~~~ \sim {M \over R^3}
  \times {M \over R c^2},~~~ \sim {M a_0^2 \over R^5} \times {M \over a_0 
    c^2},~{\rm and}~ \sim {M a_0^2 \over R^5} \times {M \over R c^2}.~~~~~
\eeqa
Here we have used the relation $\bI_{11} \sim M a_0^2$. We will derive
the four types of terms given above.

Since we consider an incompressible fluid, the gravitational potentials
inside each star are expressed as polynomials in the coordinates $x_i$.
For the purpose of obtaining the 1PN corrections given above, we need to
take into account the coefficients of terms such as $x_1^{m_1} x_2^{m_2}
x_3^{m_3}$ in the gravitational potentials of 1PN order, where $0 \le
m_1,~m_2,~m_3 \le 5$ and $0 \le m_1+m_2+m_3 ~(\equiv m_t) \le 5$ up to
$O(R^{-5})$ for the case in which $m_t$ is odd and up to $O(R^{-3})$ for
the case in which $m_t$ is even.

\subsection{Hydrostatic and Poisson equations for 1PN irrotational
  binary systems}

For an irrotational fluid, the relativistic Euler equation can be
integrated, and in the 1PN case, it is written as\cite{Shibata98}
\beqa
  {\rm constant} &=&{P \over \r} -U +{1 \over 2} \sum_k (\p_k \phi_{\rm
    N})^2 -\sum_k \ell^k \p_k \phi_{\rm N} \nonumber \\
  &&+ {1 \over c^2} \biggl[ -{P \over \r} U +{1 \over 2} U^2 +X -{1
    \over 2} \Bigl( {P \over \r} +3U \Bigr) \sum_k (\p_k \phi_{\rm N})^2 
  -{1 \over 8} \Bigl( \sum_k (\p_k \phi_{\rm N})^2 \Bigr)^2 \nonumber \\
  & &\hspace{20pt}+\sum_k (\p_k \phi_{\rm N}) (\p_k \phi_{\rm PN})
  -\sum_k \ell^k \p_k \phi_{\rm PN} -\sum_k \hat{\b}_k \p_k \phi_{\rm N}
  \biggr], \label{integEuler}
\eeqa
where we have assumed that the fluid is incompressible, i.e., $\r ={\rm
  const}$. In Eq. (\ref{integEuler}), $P$, $\r$, $\ell^k$, $\phi_{\rm
  N}$, $\phi_{\rm PN}$, $U$, $X$, and $\hat{\b}_k$ denote the pressure,
the density, the velocity field of the orbital motion (spatial component
of the timelike Killing vector), the Newtonian and 1PN velocity
potentials, and the last three terms are the Newtonian and 1PN
potentials, which are derived by solving the Poisson equations as
\beqa
  &&\D U = -4\pi \r, \\
  &&\D X = 4\pi \r \Bigl[ 2U +{3P \over \r} +2\sum_k (\p_k \phi_{\rm N})^2 
  \Bigr], \\
  &&\D P_k = -4\pi \r \p_k \phi_{\rm N}, \\
  &&\D \chi = 4\pi \r \sum_k (\p_k \phi_{\rm N}) X^k.
\eeqa
Here $\hat{\b}_k$ is expressed as
\beqa
  \hat{\b}_k =-{7 \over 2} P_k +{1 \over 2} \Bigl( \p_k \chi +\sum_l X^l 
  \p_k P_l \Bigr).
\eeqa
We note that using the gravitational potentials, the spacetime line
element to 1PN order can be written as
\beqa
  ds^2 =-\a^2 c^2 dt^2 +{2 \over c^2} \sum_i \hat{\b}_i dx^i dt +\Bigl(
  1+{2U \over c^2} \Bigr) \sum_i dx^i dx^i,
\eeqa
where
\beqa
  \a =1 -{U \over c^2} +{1 \over c^4} \Bigl( {U^2 \over 2} +X \Bigr)
  +O(c^{-6}).
\eeqa

A characteristic feature of the irrotational fluid is that the
continuity equation reduces to a Poisson-type equation for a velocity
potential $\phi$,\cite{Shibata98,Teukolsky} and in the 1PN incompressible
case, it becomes
\beqa
  \sum_i \r \p_i C_i =0,
\eeqa
where
\beqa
  C_i =-\ell^i +\p_i \phi_{\rm N} +{1 \over c^2} \Bigl[ -\ell^i \Bigl(
  {1 \over 2} \sum_k (\p_k \phi_{\rm N})^2 +3U \Bigr) -{P \over \r} \p_i 
  \phi_{\rm N} -\hat{\b}_i +\p_i  \phi_{\rm PN} \Bigr].~~~~~~~~~~
  \label{1PNintvelo}
\eeqa
In the following subsections, we obtain the terms appearing in
Eq. (\ref{integEuler}) by solving equations separately.

\subsection{Newtonian terms}

Each gravitational potential is composed of two parts. One of them is
the contribution from star 1 and the other is from star 2. In the
following, we denote the former part as $\Psi^{1 \ra 1}$ and the latter 
one as $\Psi^{2 \ra 1}$, where $\Psi$ denotes one of the potentials. We
also define $\Psi^{1 \ra 2}$ and $\Psi^{2 \ra 2}$ as the contribution
from star 1 to 2 and star 2 to itself, respectively.

In analogy to previous studies,\cite{chandra69,Aizenman,LRS94} the
configuration of each star in the binary system at Newtonian order
is assumed to be an ellipsoidal figure of its axial length $a_1$, $a_2$,
and $a_3$. In this case, the solution of the Poisson equation for the
Newtonian gravitational potential
\beqa
  \D U=-4 \pi \r,
\eeqa
is written as $U= U^{1 \ra 1} +U^{2 \ra 1}$, where
\beqa
  U^{1 \ra 1} &=& \pi \r (A_0 -\sum_l A_l x_l^2), \\
  U^{2 \ra 1} &=& {M \over R} \biggl[ 1 -{x_1 \over R} +{2x_1^2 -x_2^2
    -x_3^2 \over 2R^2} +{-2x_1^3 +3x_1 (x_2^2 +x_3^2) \over 2R^3}
  \nonumber \\
  &&\hspace{20pt}+{8x_1^4 +3x_2^4 +3x_3^4 -24x_1^2 (x_2^2 +x_3^2)
    +6x_2^2 x_3^2 \over 8R^4} \biggr] \nonumber \\
  &&+{3 \bI_{11} \over 2R^3} \biggl( 1-{3x_1 \over R}
  +{12x_1^2 -5x_2^2 -5x_3^2 \over 2R^2} \biggr) +{3 \over 2R^5}
  (\bI_{22} x_2^2 +\bI_{33} x_3^2) \nonumber \\
  & &+{1 \over 8R^5} (8I_{1111} +3I_{2222}
  +3I_{3333} -24I_{1122} -24I_{1133} +6I_{2233}) \nonumber \\
  &&+O(R^{-6}),
\eeqa
and
\beqa
  I_{iijj} =\int d^3 x \r x_i^2 x_j^2 ={M \over 35} a_i^2 a_j^2 (1+2
    \d_{ij}).
\eeqa
Here, $A_{ij \cdots}$ are index symbols introduced by
    Chandrasekhar\cite{chandra69} and $A_0=\sum_l A_l a_l^2$ is
    calculated from\cite{chandra69}
\beqa
  A_0 &=& a_1 a_2 a_3 \int_0^{\infty} {du \over \sqrt{(a_1^2 +u) (a_2^2
    +u) (a_3^2 +u)}} \nonumber \\
  &=&a_1^2 \a_2 \a_3 \int_0^{\infty} {dt \over \sqrt{(1+t) (\a_2^2 +t)
    (\a_3^2 +t)}},
\eeqa
where $\a_2=a_2/a_1$ and $\a_3=a_3/a_1$. Note that $U^{2 \ra 2}$ and
    $U^{1 \ra 2}$ are obtained by changing $x_1$ to $-(x_1 +R)$ in
    $U^{1 \ra 1}$ and $U^{2 \ra 1}$, respectively.

The pressure at Newtonian order is written as
\beqa
  P &=& P_0 \Bigl( 1- \sum_l {x_l^2 \over a_l^2} \Bigr),
\eeqa
where $P_0$ denotes the pressure at the center at the star and is
calculated from the scalar virial relation as
\beqa
  {P_0 \over \r} &=&{1 \over 3} \pi \r A_0 -{5 \over 2R^3}
  \bI_{11} -{1 \over 6} F_a (a_1^2 -a_2^2) \O_{\rm DR}^2 +O(R^{-5}).
\eeqa
The axial ratios $\a_2$ and $\a_3$ are determined from\cite{LRS94}
\beqa
  a_1^2 A_1 &=&{P_0 \over \pi \r^2} +{M a_1^2 \over \pi \r R^3} +{a_1^2
    \over 2\pi \r} \O_{\rm DR}^2 F_a (2-F_a) +O(R^{-5}), \nonumber \\
  a_2^2 A_2 &=&{P_0 \over \pi \r^2} -{M a_2^2 \over 2\pi \r R^3} -{a_2^2 
    \over 2\pi \r} \O_{\rm DR}^2 F_a (2+F_a) +O(R^{-5}), \nonumber \\
  a_3^2 A_3 &=&{P_0 \over \pi \r^2} -{M a_3^2 \over 2\pi \r R^3}
  +O(R^{-5}). \label{stvequation}
\eeqa
We can determine the equilibrium sequence of the irrotational
Darwin-Riemann ellipsoid from Eqs. (\ref{Nomega}) and
(\ref{stvequation}).

The velocity field of Newtonian order of star 1 in the inertial frame is
written as $v_i \equiv \p_i \phi_{\rm N}$, where
\beqa
  v_1 &=& \ell_1 +u_1 =-\Bigl( {a_1^2 f_R \over a_1^2 +a_2^2} +1 \Bigr)
  \O x_2, \nonumber \\
  v_2 &=& \ell_2 +u_2 =\Bigl( {a_2^2 f_R \over a_1^2 +a_2^2} +1 \Bigr)
  \O x_1 +{R \over 2} \O, \nonumber \\
  v_3 &=&0.
\eeqa
The quantity $u_i$ is the internal velocity of the star in the
corotating frame $x_i$, and $\O$ denotes the angular velocity of the
binary system. $\ell_i$ and $u_i$ are given by
\beqa
  \ell_i &=& \Bigl( -\O x_2,~\O \Bigl( x_1 +{R \over 2} \Bigr),~0
  \Bigr), \\
  u_i &=& \Bigl( {a_1 \over a_2} \L x_2,~-{a_2 \over a_1} \L x_1,~0
  \Bigr) \nonumber \\
  &=&\Bigl( -{a_1^2 f_R \over a_1^2 +a_2^2} \O x_2,~ {a_2^2 f_R \over
    a_1^2 +a_2^2} \O x_1,~ 0 \Bigr),
\eeqa
where $\L$ is the angular velocity of the internal motion. $f_R$ is
defined as $\zeta/\O$, where $\zeta \equiv ({\rm rot} {\bf u})_3$
denotes the vorticity in the corotating frame. For the irrotational
case, $f_R$ becomes $-2$. Thus, the velocity potential at Newtonian
order becomes
\beqa
  \phi_{\rm N} =F_a \O x_1 x_2 +{R \over 2} \O x_2
\eeqa
and the velocity field is written as
\beqa
  v_i &=& \Bigl( F_a \O x_2,~ F_a \O x_1 +{R \over 2} \O,~0 \Bigr),
\eeqa
where we define
\beqa
  F_a \equiv {a_1^2 -a_2^2 \over a_1^2 +a_2^2}.
\eeqa

\subsection{1PN terms}

\subsubsection{$X$}

For computational convenience, we separate $X$ into two parts
as
\beqa
  X=X_0 +X_v,
\eeqa
where $X_0$ and $X_v$ are derived from the Poisson-like equations
\beqa
  \D X_0 &=& 4\pi \r \Bigl( 2U +{3P \over \r} \Bigr), \\
  \D X_v &=& 8\pi \r \sum_k (\p_k \phi_k)^2.
\eeqa
As in the case of $U$, the 1PN potentials are divided into two parts as
$X_0 = X_0^{1 \ra 1} +X_0^{2 \ra 1}$, and we consider them separately.

The contribution from star 1 is derived from the Poisson-like equation
\beqa
  \D X_0^{1 \ra 1} =4\pi \r \Bigl[ \Bigl( 2\pi \r A_0 +{3P_0 \over \r}
  \Bigr) -\sum_l \Bigl( 2\pi \r A_l +{3P_0 \over \r a_l^2} \Bigr) x_l^2
  +2U^{2 \ra 1} \Bigr],~~~~~
\eeqa
and the solution is
\beqa
  X_0^{1 \ra 1} &=& -\a_0 U^{1 \ra 1} +\a_1 D_1 +\sum_l \eta_l D_{ll}
  -{M \over R^3} (2D_{11} -D_{22} -D_{33}) \nonumber \\
  &&+{M \over R^4} (2 D_{111} -3D_{122} -3D_{133}) +O(R^{-5}),
\eeqa
where
\beqa
  \a_0 &=&2\pi \r A_0 +{3P_0 \over \r} +{2M \over R} +{3\bI_{11} \over
    R^3} +O(R^{-5}), \\
  \a_1 &=&{2M \over R^2} +{9\bI_{11} \over R^4} +O(R^{-6}), \\
  \eta_l &=&2\pi \r A_l +{3P_0 \over \r a_l^2}.
\eeqa
Here, $D_{ij \cdots}$ are the solutions of equations
\beqa
  \D D_{ij \cdots} = -4\pi \r x_i x_j \cdots,
\eeqa
and the solutions are written as\cite{chandra69}
\beqa
  D_i &=& \pi \r a_i^2 \bigl( A_i -\sum_l A_{il} x_l^2 \bigr) x_i, \\
  D_{ii} &=& \pi \r \Bigl[ a_i^4 \bigl( A_{ii} -\sum_l A_{iil} x_l^2
  \bigr) x_i^2 \nonumber \\
  &&\hspace{15pt}+{1 \over 4} a_i^2 \bigl( B_i -2\sum_l B_{il} x_l^2
  +\sum_l \sum_m B_{ilm} x_l^2 x_m^2 \bigr) \Bigr], \\
  D_{ij} &=& \pi \r a_i^2 a_j^2 \bigl( A_{ij} -\sum_l A_{ijl} x_l^2
  \bigr) x_i x_j,~~~(i \ne j) \\
  D_{iii} &=& \pi \r \Bigl[ a_i^6 \bigl( A_{iii} -\sum_l A_{iiil} x_l^2
  \bigr) x_i^3 \nonumber \\
  &&\hspace{15pt}+{3 \over 4} a_i^4 \bigl( B_{ii} -2\sum_l B_{iil} x_l^2
  +\sum_l \sum_m B_{iilm} x_l^2 x_m^2 \bigr) x_i \Bigr], \\
  D_{ijj} &=&\pi \r \Bigl[ a_i^2 a_j^4 \bigl( A_{ijj} -\sum_l A_{ijjl}
  x_l^2 \bigr) x_i x_j^2 \nonumber \\
  &&\hspace{15pt}+{1 \over 4} a_i^2 a_j^2 \bigl( B_{ij} -2\sum_l 
  B_{ijl} x_l^2 +\sum_l \sum_m B_{ijlm} x_l^2 x_m^2 \bigr) x_i
  \Bigr],~~~(i \ne j)~~~~~~~~
\eeqa
where $B_{ijk \cdots}$ are index symbols defined by
Chandrasekhar.\cite{chandra69}

The contribution from star 2 for $X_0$ is calculated from the equation
\beqa
  \D X_0^{2 \ra 1} =4\pi \r \Bigl[ \Bigl( 2\pi \r A_0 +{3P_0 \over \r}
  \Bigr) -\sum_l \Bigl( 2\pi \r A_l +{3P_0 \over \r a_l^2} \Bigr) y_l^2
  +2U^{1 \ra 2} \Bigr],~~~~~
\eeqa
where $y_1 =-(x_1 +R)$, $y_2 =x_2$, and $y_3 =x_3$. The solution is
written
\beqa
  X_0^{2 \ra 1} &=& -\a_0 U^{2 \ra 1} -\a_1 D_1^{2 \ra 1} +\sum_l \eta_l
  D_{ll}^{2 \ra 1} \nonumber \\
  &&-{M \over R^3} (2D_{11}^{2 \ra 1} -D_{22}^{2 \ra 1}
  -D_{33}^{2 \ra 1}) +O(R^{-6}),
\eeqa
where $D_{ij \cdots}^{2 \ra 1}$ are calculated from the same equations
as in the case of $D_{ij \cdots}$, i.e.,
\beqa
  \D D_{ij \cdots}^{2 \ra 1} =-4\pi \r y_i y_j \cdots.
\eeqa
The solutions are
\beqa
  D_1^{2 \ra 1} &=& {I_{11} \over R^2} \Bigl[ 1-{2x_1 \over R} +{3 \over 
    2R^2} (2x_1^2 -x_2^2 -x_3^2) -{2 \over R^3} (2x_1^3 -3x_1 x_2^2
  -3x_1 x_3^2) \Bigr] \nonumber \\
  & &+{1 \over 2R^4} (2I_{1111} -3I_{1122} -3I_{1133})
  \Bigl( 1-{4x_1 \over R} \Bigr) +O(R^{-6}), \\
  D_i^{2 \ra 1} &=& {I_{ii} \over R^3} x_i \Bigl( 1 -{3x_1
    \over R} \Bigr) +O(R^{-5}),~~~(i \ne 1) \\
  D_{ii}^{2 \ra 1} &=& {I_{ii} \over R} \Bigl[ 1-{x_1 \over R} +{2x_1^2
    -x_2^2 -x_3^2 \over 2R^2} +{-2x_1^3 +3x_1 (x_2^2 +x_3^2) \over 2R^3} 
  \Bigr] \nonumber \\
  & &+{3\bI_{ii11} \over 2R^3} \Bigl( 1-{3x_1 \over R} \Bigr)
    +O(R^{-5}), \\
  D_{1i}^{2 \ra 1} &=& 3I_{11ii} {x_i \over R^4} \Bigl( 1 -{4x_1 \over 
    R} \Bigr) +O(R^{-6}),~~~ (i \ne 1) \\
  D_{1jj}^{2 \ra 1} &=& {I_{11jj} \over R^2} \Bigl[ 1-{2x_1 \over R} +{3 
    \over 2R^2} (2x_1^2 -x_2^2 -x_3^2) \Bigr] +O(R^{-5}), \\
  D_{2jj}^{2 \ra 1} &=& {I_{22jj} \over R^2} \Bigl[ {x_2 \over R} -{3x_1 
    x_2 \over R^2} \Bigr] +O(R^{-5}), \\
  D_{3jj}^{2 \ra 1} &=& {I_{33jj} \over R^2} \Bigl[ {x_3 \over R} -{3x_1
  x_3 \over R^2} \Bigr] +O(R^{-5}),
\eeqa
where
\beqa
  \bI_{ii11} =I_{ii11} -{1 \over 3} \sum_l I_{iill}.
\eeqa

As in the case of $X_0$, it is convenient to separate $X_v$ as
$X_v^{1 \ra 1} +X_v^{2 \ra 1}$.
Here, $X_v^{1 \ra 1}$ is derived from the Poisson-like equation
\beqa
  \D X_v^{1 \ra 1} =8 \pi \r \O^2 \Bigl[ F_a^2 (x_1^2 +x_2^2) +F_a R
  x_1 +{R^2 \over 4} \Bigr],
\eeqa
and the solution is written
\beqa
  X_v^{1 \ra 1} =-2 \O^2 \Bigl[ F_a^2 (D_{11} +D_{22}) +F_a R D_1 +{R^2
    \over 4} U^{1 \ra 1} \Bigr].
\eeqa
The equation for $X_v^{2 \ra 1}$ is
\beqa
  \D X_v^{2 \ra 1} =8 \pi \r \O^2 \Bigl[ F_a^2 (y_1^2 +y_2^2) -F_a R
  y_1 +{R^2 \over 4} \Bigr].
\eeqa
Then, the solution is found to be
\beqa
  X_v^{2 \ra 1} =-2 \O^2 \Bigl[ F_a^2 (D_{11}^{2 \ra 1} +D_{22}^{2 \ra
    1}) -F_a R D_1^{2 \ra 1} +{R^2 \over 4} U^{2 \ra 1} \Bigr].
\eeqa

\subsubsection{$\hat{\b}_k$}

Substituting $v_i$ of Newtonian order into the equations for $P_i$ and
$\chi$, we immediately find the solutions
\beqa
  P_1^{1 \ra 1} &=& F_a \O D_2, \\
  P_2^{1 \ra 1} &=& \O \Bigl( F_a D_1 +{R \over 2} U^{1 \ra 1} \Bigr),
  \\
  P_3^{1 \ra 1} &=&0, \\
  P_1^{2 \ra 1} &=& F_a \O D_2^{2 \ra 1}, \\
  P_2^{2 \ra 1} &=& \O \Bigl( F_a D_1^{2 \ra 1} -{R \over 2} U^{2 \ra 1} 
  \Bigr), \\
  P_3^{2 \ra 1} &=&0, \\
  \chi^{1 \ra 1} &=&-\O \Bigl[ 2F_a D_{12} +{R \over 2} (F_a +1) D_2
  \Bigr], \\
  \chi^{2 \ra 1} &=& -\O \Bigl[ 2F_a D_{12}^{2 \ra 1} -{R \over 2} (F_a +1)
  D_2^{2 \ra 1} \Bigr].
\eeqa

Using the solutions of $P_k$ and $\chi$ we have derived above,
we obtain
\beqa
  \hat{\b}_k \equiv \hat{\b}_k^{1 \ra 1} +\hat{\b}_k^{2 \ra 1},
\eeqa
where
\beqa
  \hat{\b}_1^{1 \ra 1} &=& {F_a \over 2} \pi \r \O x_2 \Bigl[ a_1^2 A_1
  -7a_2^2 A_2 -2a_1^2 a_2^2 A_{12} \nonumber \\
  & &\hspace{50pt}+(5a_2^2 A_{12} -3a_1^2 A_{11}
  +6a_1^2 a_2^2 A_{112}) x_1^2 \nonumber \\
  & &\hspace{50pt}+(7a_2^2 A_{22} -a_1^2 A_{12} +2a_1^2 
  a_2^2 A_{122}) x_2^2 \nonumber \\
  & &\hspace{50pt}+(7a_2^2 A_{23} -a_1^2 A_{13} +2a_1^2 a_2^2
  A_{123}) x_3^2 \Bigl] \nonumber \\
  & &-{R \over 2} \pi \r \O (A_1 -a_2^2 A_{12}) x_1 x_2, \\
  \hat{\b}_2^{1 \ra 1} &=& {F_a \over 2} \pi \r \O x_1 \Bigl[
  a_2^2 A_2 -7a_1^2 A_1 -2a_1^2 a_2^2 A_{12} \nonumber \\
  & &\hspace{50pt}+(7a_1^2 A_{11} -a_2^2
    A_{12} +2a_1^2 a_2^2 A_{112}) x_1^2 \nonumber \\
  & &\hspace{50pt}+(5a_1^2 A_{12} -3a_2^2 A_{22} +6a_1^2 a_2^2 A_{122})
  x_2^2 \nonumber \\
  & &\hspace{50pt}+(7a_1^2 A_{13} -a_2^2 A_{23} +2a_1^2 a_2^2 A_{123}) x_3^2
  \Bigr] \nonumber \\
  & &+{R \over 4} \pi \r \O \Bigl[ -7A_0 -a_2^2 A_2 +(7A_1 +
  a_2^2 A_{12}) x_1^2 +(5A_2 +3 a_2^2 A_{22}) x_2^2 \nonumber \\
  &&\hspace{50pt}+(7A_3 +a_2^2 A_{23}) x_3^2 \Bigr], \\
  \hat{\b}_3^{1 \ra 1} &=& -{1 \over 2} \pi \r \O x_2 x_3 \Bigl[ 2F_a
  (a_1^2 A_{13} +a_2^2 A_{23} -2a_1^2 a_2^2 A_{123}) x_1 \nonumber \\
  &&\hspace{65pt}+R (A_3 -a_2^2 A_{23}) \Bigr], \\
  \hat{\b}_1^{2 \ra 1} &=& {\O \over 2} \Bigl[ {M x_2 \over
    2R} -{x_2 \over R^3} \Bigl( 2F_a I_{11} +10F_a I_{22} +{3 \over 2}
  I_{22} -{9 \over 4} \bI_{11} \Bigr) -{M \over R^2} x_1 x_2
  \nonumber \\
  &&\hspace{20pt}+{3M \over 4R^3} x_2 (2x_1^2 -x_2^2 -x_3^2) +O(R^{-4})
    \Bigr], \\
  \hat{\b}_2^{2 \ra 1} &=& {\O \over 2} \Bigl[ {7M \over 2} +{1 \over
    R^2} \Bigl( -7F_a I_{11} +F_a I_{22} +{I_{22} \over 2} +{21 \bI_{11}
    \over 4} \Bigr) \nonumber \\
  & &\hspace{20pt}+{x_1 \over R} \Bigl\{ -{7M \over 2} +{1 \over R^2}
  \Bigl( 14F_a I_{11} -I_{22} \Bigl( 2F_a +{3 \over 2} \Bigr) -{63
    \bI_{11} \over 4} \Bigr) \Bigr\} \nonumber \\
  & &\hspace{20pt}+{7M \over 4R^2}
    (2x_1^2 -x_2^2 -x_3^2) +{M \over 2R^2} x_2^2
    -{7M \over 4R^3} x_1 (2x_1^2 -3x_2^2 -3x_3^2) \nonumber \\
  &&\hspace{20pt}-{3M \over 2R^3} x_1 x_2^2 +O(R^{-4}) \biggr], \\
  \hat{\b}_3^{2 \ra 1} &=& {M\O \over 4R^2} x_2 x_3 \biggl[ 1 -{3x_1 \over
    R} +O(R^{-2}) \Bigr].
\eeqa
Then, the divergence of $\hat{\b}_k$ is
\beqa
  \sum_k \p_k \hat{\b}_k =\sum_k \p_k (\hat{\b}_k^{1 \ra 1}
  +\hat{\b}_k^{2 \ra 1}),
\eeqa
where
\beqa
  \sum_k \p_k \hat{\b}_k^{1 \ra 1} &=&3\pi \r \O \Bigl[ 2F_a (a_1^2
    +a_2^2) A_{12} x_1 x_2 +RA_2 x_2 \Bigr], \\
  \sum_k \p_k \hat{\b}_k^{2 \ra 1} &=&-{3 M \over 2R^2} \O x_2
  \Bigl( 1-{3x_1 \over R} +O(R^{-2}) \Bigr).
\eeqa
Here we have used some relations among index symbols.\cite{chandra69}

\subsubsection{$\phi_{\rm PN}$}

An explicit form of the Poisson-like equation for $\phi_{\rm PN}$ can be 
written as
\beqa
  \D \phi_{\rm PN} &=& \p_i \Bigl[ \ell^i \Bigl( {1 \over 2} \sum_k (\p_k
  \phi_{\rm N})^2 +3U \Bigr) +{P \over \r} \p_i \phi_{\rm N} +\hat{\b}_i 
  \Bigr], \nonumber \\
  &=&-{2P_0 \over \r} {a_1^2 -a_2^2 \over a_1^2 a_2^2} \O x_1 x_2 -R \O 
  \Bigl[ {a_2^2 (a_1^2 -a_2^2) \over (a_1^2 +a_2^2)^2} \O^2 +{P_0 \over
    \r a_2^2} \Bigr] x_2 \nonumber \\
  &&+\O \times O(R^{-4}).
\eeqa
The solution of this equation can be written up to biquadratic terms in
$x_i$ as\footnote{Although we can add higher order terms which satisfy
  $\D \phi_{\rm PN}=0$, we neglect them for simplicity because we
  consider only biquadratic deformation of ellipsoids in this paper.}
\beqa
  \phi_{\rm PN} =(p+qx_1^2 +rx_2^2 +sx_3^2) x_1 x_2 +(e+fx_1^2 +gx_2^2
  +hx_3^2) x_2 +{\rm const},~~~~~~~~\label{1PNvelopot}
\eeqa
where $q$, $r$, $s$, $f$, $g$, and $h$ satisfy the conditions
\beqa
  3q+3r+s &=& -{P_0 \over \r} {a_1^2 -a_2^2 \over a_1^2 a_2^2} \O,
  \label{cond1} \\
  f+3g+h &=& -{R \over 2} \O \Bigl[ {a_2^2 (a_1^2 -a_2^2) \over (a_1^2
    +a_2^2)^2} \O^2 +{P_0 \over \r a_2^2} \Bigr]. \label{cond2}
\eeqa
On the right-hand side of Eq. (\ref{1PNvelopot}), coefficients $p$, $q$,
    $r$, and $s$ are concerned with the star itself. We can find these
    coefficients in a previous paper\cite{TAS} in which we
    investigated irrotational and incompressible stars. On the other
    hand, the coefficients $e$, $f$, $g$, and $h$ are associated with the
    binary motion.

\subsection{Collection}

Substituting the terms derived in the previous subsections into
Eq. (\ref{integEuler}), we obtain
\beqa
  {\cal G} &\equiv&{P \over \r} -U -\d U \nonumber \\
  & &-{1 \over 2} \Bigl( \O_{\rm
    DR}^2 +{1 \over c^2} \d \O^2 \Bigr) \Bigl[ F_a (2-F_a) x_1^2 -F_a
  (2+F_a) x_2^2 +R x_1 +{R^2 \over 4} \Bigr] \nonumber \\
  & &+{1 \over c^2} \Bigl[ \g_0 +\sum_l \g_l x_l^2 +\sum_{l \le m}
  \g_{lm} x_l^2 x_m^2 +x_1 \bigl( \k_0 +\sum_l \k_l x_l^2 +\sum_{l \le
    m} \k_{lm} x_l^2 x_m^2 \bigr) \Bigr] \nonumber \\
  &=&{\rm const}, \label{collecEuler}
\eeqa
where $\O_{\rm DR}$ denotes the angular velocity of an irrotational
Darwin-Riemann ellipsoid, and $\d \O$ denotes the 1PN correction for the
angular velocity.  $\g_{ij}$ and $\k_{ij}$ are expressed as
\beqa
  \g_0 &=&\pi \r \biggl[ {1 \over 2} \pi \r A_0^2 -{P_0 \over \r} A_0
    -\Bigl( {\a_0 \over \pi \r} +{P_0 \over \pi \r^2} -A_0 \Bigr) \Bigl(
    {M \over R} +{3\bI_{11} \over 2R^3} \Bigr) -\a_0 A_0 \nonumber \\
  &&\hspace{20pt}+{1 \over 4}
    \sum_l \eta_l a_l^2 B_l +{1 \over \pi \r} \sum_l \eta_l \Bigl(
    {I_{ll} \over R} +{3\bI_{ll11} \over 2R^3} \Bigr) \nonumber \\
  & &\hspace{20pt}-{M \over 4R^3} (2a_1^2 B_1 -a_2^2 B_2 -a_3^2 B_3) -{1
    \over 2} \O^2 F_a^2 (a_1^2 B_1 +a_2^2 B_2) \nonumber \\
  & &\hspace{20pt}-{1 \over 8} \O^2 R^2 \Bigl( {P_0 \over \pi \r^2}
    -a_2^2 A_2 \Bigr) +O(R^{-5}) \biggr] \nonumber \\
  & &+{M \over 2R^2} \Bigl( M-{3\bI_{11} \over R^2} \Bigr)
  -\a_1 {I_{11} \over R^2} -{R^4 \over 128} \O^4 \nonumber \\
  & &-{\O^2 \over 8} \Bigl[ 14MR +{1 \over R} \Bigl\{ 16F_a^2 (I_{11}
  +I_{22}) -30F_a I_{11} + (2F_a +1) I_{22} +21\bI_{11} \Bigr\} \Bigr]
  \nonumber \\
  &&+O(R^{-6}), \\
  \g_1 &=& \pi \r \biggl[ (\a_0 -\pi \r A_0 ) A_1 +{P_0 \over \r a_1^2}
    (A_0 +a_1^2 A_1) -{1 \over a_1^2} \Bigl( a_1^2 A_1 -{P_0 \over \pi
      \r^2} \Bigr) \Bigl( {M \over R} +{3\bI_{11} \over 2R^3} \Bigr)
    \nonumber \\
  &&\hspace{20pt}-\Bigl( {\a_0 \over \pi \r} +{P_0 \over \pi \r^2} -A_0
    \Bigr) {M \over R^3} +\eta_1 a_1^2 \Bigl( a_1^2 A_{11} -{1 \over 2}
    B_{11} \Bigr) -{1 \over 2} \eta_2 a_2^2 B_{12} \nonumber \\
  & &\hspace{20pt}-{1 \over 2} \eta_3 a_3^2 B_{13}
    +{1 \over \pi \r R^3} \sum_l \eta_l I_{ll} \nonumber \\
  & &\hspace{20pt}-{M \over 2R^3} (4a_1^4
    A_{11} -2a_1^2 B_{11} +a_2^2 B_{12} +a_3^2 B_{13}) \nonumber \\
  & &\hspace{20pt}-{1 \over 2} \O^2 
    \Bigl\{ {R^2 \over 4} \Bigl( a_2^2 A_{12} -{P_0 \over \pi \r^2
      a_1^2} \Bigr) +F_a^2 \Bigl( 3A_0 -7a_1^2 A_1 +a_2^2 A_2 \nonumber \\
  &&\hspace{70pt}+4a_1^4
    A_{11} -2a_1^2 a_2^2 A_{12} -2a_1^2 B_{11} -2a_2^2 B_{12} +{P_0
      \over \pi \r^2} \Bigr) \Bigr\} \biggr] \nonumber \\
  &&+(F_a-1) \O p +O(R^{-4}), \\
  \g_2 &=&\pi \r \biggl[ (\a_0 -\pi \r A_0) A_2 +{P_0 \over \r a_2^2}
    (A_0 +a_2^2 A_2) -{1 \over a_2^2} \Bigl( a_2^2 A_2 -{P_0 \over \pi
      \r^2} \Bigr) \Bigl( {M \over R} +{3\bI_{11} \over 2R^3} \Bigr)
    \nonumber \\
  &&\hspace{20pt}+\Bigl( {\a_0 \over \pi \r} +{P_0 \over \pi \r^2} -A_0
    \Bigr) {M \over 2R^3} -{1 \over 2}\eta_1 a_1^2 B_{12} +\eta_2 a_2^2
    \Bigl( a_2^2 A_{22} -{1 \over 2} B_{22} \Bigr) \nonumber \\
  & &\hspace{20pt}-{1 \over 2} \eta_3 a_3^2
    B_{23} -{1 \over 2 \pi \r R^3} \sum_l \eta_l I_{ll} \nonumber \\
  & &\hspace{20pt}+{M \over 2R^3}
    (2a_2^4 A_{22} +2a_1^2 B_{12} -a_2^2 B_{22} -a_3^2 B_{23}) \nonumber \\
  & &\hspace{20pt}-{1 \over 2} \O^2 \Bigl\{ {R^2 \over 4} \Bigl( 3a_2^2
    A_{22} -2A_2 -{P_0 \over \pi \r^2 a_2^2} \Bigr) +F_a^2 \Bigl( 3A_0
    +a_1^2 A_1 -7a_2^2 A_2 \nonumber \\
  & &\hspace{70pt}+4a_2^4 A_{22} -2a_1^2 a_2^2 A_{12} -2a_1^2
    B_{12} -2a_2^2 B_{22} +{P_0 \over \pi \r^2} \Bigr) \Bigr\}
    \biggr] \nonumber \\
  &&+(F_a +1) \O p +O(R^{-4}), \\
  \g_3 &=& \pi \r \biggl[ (\a_0 -\pi \r A_0) A_3 +{P_0 \over \r a_3^2}
    (A_0 +a_3^2 A_3) -{1 \over a_3^2} \Bigl( a_3^2 A_3 -{P_0 \over \pi
      \r^2} \Bigr) \Bigl( {M \over R} +{3\bI_{11} \over 2R^3} \Bigr)
    \nonumber \\
  &&\hspace{20pt}+\Bigl( {\a_0 \over \pi \r} +{P_0 \over \pi \r^2} -A_0
    \Bigr) {M \over 2R^3} -{1 \over 2}\eta_1 a_1^2 B_{13} -{1 \over 2}
    \eta_2 a_2^2 B_{23} \nonumber \\
  & &\hspace{20pt}+\eta_3 a_3^2 \Bigl( a_3^2 A_{33} -{1 \over 2} B_{33}
    \Bigr) -{1 \over 2 \pi \r R^3} \sum_l \eta_l I_{ll} \nonumber \\
  & &\hspace{20pt}+{M \over 2R^3}
    (2a_3^4 A_{33} +2a_1^2 B_{13} -a_2^2 B_{23} -a_3^2 B_{33}) \nonumber \\
  & &\hspace{20pt}-{1 \over 2} \O^2 \Bigl\{ {R^2 \over 4} \Bigl( a_2^2
    A_{23} -{P_0 \over \pi \r^2 a_3^2} \Bigr) -2F_a^2 (a_1^2 B_{13}
    +a_2^2 B_{23}) \Bigr\} \biggr] \nonumber \\
  &&+O(R^{-4}), \\
  \g_{11} &=&\pi \r \biggl[ -\pi \r {A_1 \over a_1^2} \Bigl( {P_0 \over \pi 
    \r^2} -{1 \over 2} a_1^2 A_1 \Bigr) -{M \over a_1^2 R^3} \Bigl( a_1^2
    A_1 -{P_0 \over \pi \r^2} \Bigr) \nonumber \\
  & &\hspace{20pt}-\eta_1 a_1^2 \Bigl( a_1^2 A_{111}
    -{1 \over 4} B_{111} \Bigr)
    +{1 \over 4} \eta_2 a_2^2 B_{112} +{1
      \over 4} \eta_3 a_3^2 B_{113} \nonumber \\
  & &\hspace{20pt}+{M \over 4R^3} (8a_1^4 A_{111}
    -2a_1^2 B_{111} +a_2^2 B_{112} +a_3^2 B_{113}) \nonumber \\
  & &\hspace{20pt}-{1 \over 2} \O^2
    F_a^2 \Bigl( -3A_1 +7a_1^2 A_{11} -a_2^2 A_{12} +2a_1^2 a_2^2
    A_{112} -4a_1^4 A_{111} \nonumber \\
  & &\hspace{70pt}+a_1^2 B_{111} +a_2^2 B_{112} -{P_0 \over
    \pi \r^2 a_1^2} \Bigr) +O(R^{-5}) \biggr] \nonumber \\
  & &+(F_a -1) \O q +O(R^{-6}), \\
  \g_{12} &=&\pi \r \biggl[ -{P_0 \over \r} \Bigl( {A_2 \over a_1^2} +{A_1
    \over a_2^2} \Bigr) +\pi \r A_1 A_2 -{M \over 2R^3} \Bigl( 2A_2 -A_1 
    -{2P_0 \over \pi \r^2 a_2^2} +{P_0 \over \pi \r^2 a_1^2} \Bigr)
    \nonumber \\
  & &\hspace{20pt}-\eta_1 a_1^2 \Bigl( a_1^2 A_{112} -{1 \over 2}
    B_{112} \Bigr) -\eta_2 a_2^2
    \Bigl( a_2^2 A_{122} -{1 \over 2} B_{122} \Bigr) +{1 \over 2} \eta_3 
    a_3^2 B_{123} \nonumber \\
  & &\hspace{20pt}+{M \over 2R^3} (4a_1^4 A_{112} -2a_2^4 A_{122}
    -2a_1^2 B_{112} +a_2^2 B_{122} +a_3^2 B_{123}) \nonumber \\
  & &\hspace{20pt}-{1 \over 2} \O^2
    F_a^2 \Bigl\{ -3(A_1 +A_2) +5(a_1^2 +a_2^2) A_{12} -3(a_1^2 A_{11}
    +a_2^2 A_{22}) \nonumber \\
  & &\hspace{70pt}+6a_1^2 a_2^2 (A_{112} +A_{122}) -4a_1^4 A_{112}
    -4a_2^4 A_{122} +2a_1^2 B_{112} \nonumber \\
  & &\hspace{70pt}+2a_2^2 B_{122} -{P_0 \over \pi \r^2} \Bigl( {1 \over
     a_1^2} +{1 \over a_2^2} \Bigr) \Bigr\} +O(R^{-5}) \biggr] \nonumber \\
  & &+3\O \bigl\{ (F_a+1) q +(F_a-1) r \bigr\} +O(R^{-6}), \\
  \g_{13} &=&\pi \r \biggl[ -{P_0 \over \r} \Bigl( {A_3 \over a_1^2} +{A_1
    \over a_3^2} \Bigr) +\pi \r A_1 A_3 -{M \over 2R^3} \Bigl( 2A_3 -A_1
    -{2P_0 \over \pi \r^2 a_3^2} +{P_0 \over \pi \r^2 a_1^2} \Bigr)
    \nonumber \\
  & &\hspace{20pt}-\eta_1 a_1^2 \Bigl( a_1^2 A_{113} -{1 \over 2}
    B_{113} \Bigr) +{1 \over 2} \eta_2 a_2^2 B_{123} -\eta_3 a_3^2 \Bigl(
    a_3^2 A_{133} -{1 \over 2} B_{133} \Bigr) \nonumber \\
  & &\hspace{20pt}+{M \over 2R^3} (4a_1^4
    A_{113} -2a_3^4 A_{133} -2a_1^2 B_{113} +a_2^2 B_{123} +a_3^2
    B_{133}) \nonumber \\
  & &\hspace{20pt}-{1 \over 2} \O^2 F_a^2 \Bigl( -3A_3 +7a_1^2 A_{13}
    -a_2^2 A_{23} +2a_1^2 a_2^2 A_{123} -4a_1^4 A_{113} \nonumber \\
  & &\hspace{70pt}+2a_1^2 B_{113}
    +2a_2^2 B_{123} -{P_0 \over \pi \r^2 a_3^2} \Bigr) +O(R^{-5})
    \biggr] \nonumber \\
  & &+(F_a -1) \O s +O(R^{-6}), \\
  \g_{22} &=& \pi \r \biggl[ -\pi \r {A_2 \over a_2^2} \Bigl( {P_0 \over
    \pi \r^2} -{1 \over 2} a_2^2 A_2 \Bigr) -{M \over 2a_2^2 R^3} \Bigl(
    {P_0 \over \pi \r^2} -a_2^2 A_2 \Bigr) \nonumber \\
  & &\hspace{20pt}+{1 \over 4} \eta_1 a_1^2
    B_{122} -\eta_2 a_2^2 \Bigl( a_2^2 A_{222} -{1 \over 4} B_{222}
    \Bigr) +{1 \over 4} \eta_3 a_3^2 B_{223} \nonumber \\
  & &\hspace{20pt}-{M \over 4R^3} (4a_2^4
    A_{222} +2a_1^2 B_{122} -a_2^2 B_{222} -a_3^2 B_{223}) \nonumber \\
  & &\hspace{20pt}-{1 \over 2}
    \O^2 F_a^2 \Bigl( -3A_2 +7a_2^2 A_{22} -a_1^2 A_{12} +2a_1^2 a_2^2
    A_{122} -4a_2^4 A_{222} \nonumber \\
  & &\hspace{70pt}+a_1^2 B_{122} +a_2^2 B_{222} -{P_0 \over \pi 
      \r^2 a_2^2} \Bigr) +O(R^{-5}) \biggr] \nonumber \\
  & &+(F_a +1) \O r +O(R^{-6}), \\
  \g_{23} &=& \pi \r \biggl[ -{P_0 \over \r} \Bigl( {A_3 \over a_2^2} +{A_2 
    \over a_3^2} \Bigr) +\pi \r A_2 A_3 -{M \over 2R^3} \Bigl( -A_2 -A_3 
    +{P_0 \over \pi \r^2 a_2^2} +{P_0 \over \pi \r^2 a_3^2} \Bigr)
    \nonumber \\
  & &\hspace{20pt}+{1
      \over 2} \eta_1 a_1^2 B_{123} -\eta_2 a_2^2 \Bigl( a_2^2 A_{223}
    -{1 \over 2} B_{223} \Bigr) \nonumber \\
  & &\hspace{20pt}-\eta_3 a_3^2 \Bigl( a_3^2 A_{233} -{1
      \over 2} B_{233} \Bigr) \nonumber \\
  &&\hspace{20pt}-{M \over 2R^3} (2a_2^4 A_{223} +2a_3^4
    A_{233} +2a_1^2 B_{123} -a_2^2 B_{223} -a_3^2 B_{233}) \nonumber \\
  & &\hspace{20pt}-{1 \over 2}
    \O^2 F_a^2 \Bigl( -3A_3 +7a_2^2 A_{23} -a_1^2 A_{13} +2a_1^2 a_2^2
    A_{123} -4a_2^4 A_{223} \nonumber \\
  & &\hspace{70pt}+2a_1^2 B_{123} +2a_2^2 B_{223} -{P_0 \over
      \pi \r^2 a_3^2} \Bigr) +O(R^{-5}) \biggr] \nonumber \\
  & &+(F_a+1) \O s +O(R^{-6}), \\
  \g_{33} &=&\pi \r \biggl[ -\pi \r {A_3 \over a_3^2} \Bigl( {P_0 \over \pi 
    \r^2} -{1 \over 2} a_3^2 A_3 \Bigr) -{M \over 2a_3^2 R^3} \Bigl(
    {P_0 \over \pi \r^2} -a_3^2 A_3 \Bigr) \nonumber \\
  & &\hspace{15pt}+{1 \over 4} \eta_1 a_1^2
    B_{133} +{1 \over 4} \eta_2 a_2^2 B_{233} -\eta_3 a_3^2 \Bigl( a_3^2 
    A_{333} -{1 \over 4} B_{333} \Bigr) \nonumber \\
  & &\hspace{15pt}-{M \over 4R^3} (4a_3^4 A_{333}
    +2a_1^2 B_{133} -a_2^2 B_{233} -a_3^2 B_{333}) \nonumber \\
  & &\hspace{15pt}-{1 \over 2} \O^2
    F_a^2 (a_1^2 B_{133} +a_2^2 B_{233}) +O(R^{-5}) \biggr], \\
  \k_0 &=&\pi \r \biggl[ \Bigl( {P_0 \over \pi \r^2} +{\a_0 \over \pi
    \r} -A_0 \Bigr) \Bigl( {M \over R^2} +{9\bI_{11} \over 2R^4} \Bigr)
    +\a_1 a_1^2 A_1 -\sum_l {\eta_l \over \pi \r} \Bigl( {I_{ll} \over R^2}
    +{9\bI_{ll11} \over 2R^4} \Bigr) \nonumber \\
  & &\hspace{20pt}+{3M \over 4R^4} a_1^2 (2a_1^2
    B_{11} -a_2^2 B_{12} -a_3^2 B_{13}) \nonumber \\
  & &\hspace{20pt}-{1 \over 4} \O^2 F_a R \Bigl
    ( -A_0 +a_1^2 A_1 -2a_1^2 a_2^2 A_{12} +{2P_0 \over \pi \r^2} \Bigr) 
    +O(R^{-6}) \biggr] \nonumber \\
  & &-{M^2 \over R^3} -{3M \over R^5} \bI_{11} +{2\a_1
    \over R^3} I_{11} \nonumber \\
  & &-{1 \over 8} \O^2 \Bigl[ 2(13F_a-7) M -{1 \over R^2} \Bigl\{
    4F_a (11F_a -15) I_{11} +(12F_a^2 +2F_a +3) I_{22} \nonumber \\
  & &\hspace{70pt}+3(21 -13F_a) \bI_{11} \Bigr\} \Bigr] \nonumber \\
  & &-{1 \over 16} \O^4 R^3 F_a +(F_a -1) \O e +O(R^{-7}), \\
  \k_1 &=& \pi \r \biggl[ \Bigl( {P_0 \over \pi \r^2} +{\a_0 \over \pi
    \r} -A_0 \Bigr) {M \over R^4} -{1 \over a_1^2} \Bigl( {P_0 \over \pi 
    \r^2} -a_1^2 A_1 \Bigr) \Bigl( {M \over R^2} +{9\bI_{11} \over 2R^4}
    \Bigr) \nonumber \\
  & &\hspace{20pt}-\a_1 a_1^2 A_{11} -{1 \over \pi \r R^4} \sum_l \eta_l
    I_{ll} \nonumber \\
  & &\hspace{20pt}+{M \over 2R^4} a_1^2 (4a_1^4 A_{111} -6a_1^2 B_{111}
  +3a_2^2 B_{112} +3a_3^2 B_{113}) \nonumber \\
  & &\hspace{20pt}-{1 \over 4} \O^2 R F_a \Bigl( A_1 -a_1^2 A_{11}
    +2a_1^2 a_2^2 A_{112} -{2P_0 \over \pi \r^2 a_1^2} \Bigr) +O(R^{-6}) 
    \biggr] \nonumber \\
  & &-{2M^2 \over R^5} +{M \over 4R^2} \O^2 (6F_a^2 -13F_a
    +7) -{1 \over 4} \O^4 R F_a^3 +(F_a -1) \O f \nonumber \\
  & &+O(R^{-7}), \\
  \k_2 &=& \pi \r \biggl[ -\Bigl( {P_0 \over \pi \r^2} +{\a_0 \over \pi
    \r} -A_0 \Bigr) {3M \over 2R^4} -{1 \over a_2^2} \Bigl( {P_0 \over
    \pi \r^2} -a_2^2 A_2 \Bigr) \Bigl( {M \over R^2} +{9\bI_{11} \over
    2R^4} \Bigr) \nonumber \\
  & &\hspace{20pt}-\a_1 a_1^2 A_{12} +{3 \over 2\pi \r R^4} \sum_l \eta_l 
    I_{ll} \nonumber \\
  & &\hspace{20pt}-{3M \over 2R^4} a_1^2 (2a_2^4 A_{122} +2a_1^2 B_{112}
    -a_2^2 B_{122} -a_3^2 B_{123}) \nonumber \\
  & &\hspace{20pt}+{1 \over 4} \O^2 R F_a \Bigl( 2A_1 +A_2
    +3a_1^2 A_{12} -2a_2^2 A_{12} -6a_1^2 a_2^2 A_{122} +{2P_0 \over \pi 
      \r^2 a_2^2} \Bigr) \nonumber \\
  & &\hspace{20pt}+O(R^{-6}) \biggr] \nonumber \\
  & &+{2M^2 \over R^5} +{3M \over 8R^2}
    \O^2 (4F_a^2 +5F_a -6) -{1 \over 4} \O^4 R F_a^3 +2(F_a+1) \O f
    \nonumber \\
  & &+3(F_a -1) \O g +O(R^{-7}), \\
  \k_3 &=& \pi \r \biggl[ -\Bigl( {P_0 \over \pi \r^2} +{\a_0 \over \pi
    \r} -A_0 \Bigr) {3M \over 2R^4} -{1 \over a_3^2} \Bigl( {P_0 \over
    \pi \r^2} -a_3^2 A_3 \Bigr) \Bigl( {M \over R^2} +{9\bI_{11} \over
    2R^4} \Bigr) \nonumber \\
  & &\hspace{20pt}-\a_1 a_1^2 A_{13} +{3 \over 2\pi \r R^4} \sum_l \eta_l
    I_{ll} \nonumber \\
  & &\hspace{20pt}-{3M \over 2R^4} a_1^2 (2a_3^4 A_{133} +2a_1^2 B_{113} 
    -a_2^2 B_{123} -a_3^2 B_{133})
    \nonumber \\
  & &\hspace{20pt}-{1 \over 4} \O^2 R F_a \Bigl( A_3 -a_1^2 A_{13}
    +2a_1^2 a_2^2 A_{123} -{2P_0 \over \pi \r^2 a_3^2} \Bigr) +O(R^{-6}) 
    \biggr] \nonumber \\
  & &+{2M^2 \over R^5} +{M \over 8R^2} \O^2 (13F_a -21) +(F_a -1) \O h
    +O(R^{-7}), \\
  \k_{11} &=& {M \pi \r \over R^4} \Bigl[ -{P_0 \over \pi \r^2 a_1^2} +A_1
    \nonumber \\
  & &\hspace{20pt}-{a_1^2 \over 4} (8a_1^4 A_{1111} -6a_1^2 B_{1111}
    +3a_2^2 B_{1112} +3a_3^2 B_{1113}) \Bigr], \\
  \k_{12} &=& {M \pi \r \over 2R^4} \Bigl[ -{P_0 \over \pi \r^2} \Bigl( {2
    \over a_2^2} -{3 \over a_1^2} \Bigr) -3A_1 +2A_2 \nonumber \\
  & &\hspace{20pt}-a_1^2 (4a_1^4
    A_{1112} -6a_2^4 A_{1122} -6a_1^2 B_{1112} +3a_2^2 B_{1122} +3a_3^2
    B_{1123}) \Bigr], \\
  \k_{13} &=& {M \pi \r \over 2R^4} \Bigl[ -{P_0 \over \pi \r^2} \Bigl( {2
    \over a_3^2} -{3 \over a_1^2} \Bigr) -3A_1 +2A_3 \nonumber \\
  & &\hspace{20pt}-a_1^2 (4a_1^4
    A_{1113} -6a_3^4 A_{1133} -6a_1^2 B_{1113} +3a_2^2 B_{1123} +3a_3^2
    B_{1133}) \Bigr], \\
  \k_{22} &=& {3M \pi \r \over 2R^4} \Bigl[ {P_0 \over \pi \r^2 a_2^2}
    -A_2 \nonumber \\
  & &\hspace{20pt}+{a_1^2 \over 2} (4a_2^4 A_{1222} +2a_1^2 B_{1122}
    -a_2^2 B_{1222} -a_3^2 B_{1223}) \Bigr], \\
  \k_{23} &=& {3M \pi \r \over 2R^4} \Bigl[ {P_0 \over \pi \r^2} \Bigl(
    {1 \over a_2^2} +{1 \over a_3^2} \Bigr) -A_2 -A_3 \nonumber \\
  & &\hspace{20pt}+a_1^2 (2a_2^4 A_{1223} +2a_3^4 A_{1233} +2a_1^2
    B_{1123} -a_2^2 B_{1223} -a_3^2 B_{1233}) \Bigr], \\
\k_{33} &=& {3M \pi \r \over 2R^4} \Bigl[ {P_0 \over \pi \r^2 a_3^2}
    -A_3 \nonumber \\
  & &\hspace{20pt}+{a_1^2 \over 2} (4a_3^4 A_{1333} +2a_1^2 B_{1133}
    -a_2^2 B_{1233} -a_3^2 B_{1333}) \Bigr].
\eeqa
The quantity $\d U$ denotes the gravitational potential induced by the
    deformation of the binary system. The explicit form of $\d U$ is
    given in Appendix \ref{deltaUform}.

\section{The post-Newtonian angular velocity}\label{secangvelo}

In this section, the 1PN correction for the orbital angular velocity at
1PN order is derived using the first tensor virial (TV) equation.
The first TV relation is derived from
\beqa
  \int d^3 x {\p P \over \p x_1} =0. \label{ftveq}
\eeqa
Substituting Eq. (\ref{collecEuler}) into Eq. (\ref{ftveq}), we have
Eq. (\ref{Nomega}) for Newtonian order. At 1PN order, the
explicit form becomes
\beqa
  0&=& {MR \over 2} \d \O^2 +\d \int d^3 x \r {\p U \over \p x_1}
  +\O_{\rm DR}^2 F_a (2-F_a) \int d^3 x \r \xi_1 \nonumber \\
  & &-\bigl( \k_0 M +3\k_1 I_{11} +\k_2
  I_{22} +\k_3 I_{33} +5\k_{11} I_{1111} +3\k_{12} I_{1122} +3\k_{13}
  I_{1133} \nonumber \\
  & &\hspace{15pt}+\k_{22} I_{2222} +\k_{23} I_{2233} +\k_{33} I_{3333}
  \bigr), \label{preomega}
\eeqa
where $\xi_1$ is the $x_1$ component of the Lagrangian displacement
vectors $\xi_i$.

\subsection{Deformation of the figure} \label{deform}

The density profile of an incompressible and homogeneous sphere at
1PN order is the same as that at Newtonian order. However, the
density profile of an ellipsoid at 1PN order is different from that
at Newtonian order. Chandrasekhar's method to obtain the 1PN
correction to the Newtonian figure is as follows.\cite{chandra71} First,
the ellipsoidal figures at Newtonian order are constructed. Next,
the 1PN effect is regarded as a small perturbation to the Newtonian
configuration, and the deformation from the Newtonian ellipsoid is
calculated by using the Lagrangian displacement vectors. Finally, by
solving equations for the Lagrangian displacement and calculating the
correction to $U$ by the deformation, the equilibrium configuration at
1PN order is obtained. In this paper, we follow this method.

In choosing the Lagrangian displacement vectors $\xi_k^{(ij)}$, we
require them to be divergent free $\sum_k \p_k \xi_k^{(ij)} =0$, due to
incompressibility. To obtain $\d \O^2$ up to $O(R^{-3}) \times
\O_{\rm DR}^2$, we write the Lagrangian displacement vectors as
\beqa
  \xi_k ={1 \over c^2} \sum_{ij} S_{ij} \xi_k^{(ij)},
\eeqa
where
\beqa
  \xi_k^{(11)} &=& (x_1,~0,~-x_3), \label{Lagdisp1} \\
  \xi_k^{(12)} &=& (0,~x_2,~-x_3), \\
  \xi_k^{(31)} &=& \Bigl( {1 \over 3} x_1^3,~-x_1^2 x_2,~0 \Bigr),
  \label{Lagdisp3} \\
  \xi_k^{(32)} &=& \Bigl( 0,~{1 \over 3} x_2^3,~-x_2^2 x_3 \Bigr), \\
  \xi_k^{(33)} &=& \Bigl( -x_3^2 x_1,~0,~{1 \over 3} x_3^3 \Bigr),
  \label{Lagdisp5} \\
  \xi_k^{(0)} &=& \Bigl( {1 \over 2},~0,~0 \Bigr), \label{Lagdisp6} \\
  \xi_k^{(21)} &=& \Bigl( {1 \over 2} x_1^2,~0,~-x_1 x_3 \Bigr),
  \label{Lagdisp7} \\
  \xi_k^{(22)} &=& (0,~x_1 x_2,~-x_1 x_3). \label{Lagdisp8}
\eeqa
Here, we consider only up to the biquadratic deformation so that the
Lagrangian displacement vectors of higher order functions in $x_i$ can
be neglected. The Lagrangian displacement vectors (\ref{Lagdisp1})
-- (\ref{Lagdisp5}) are concerned with the deformation of the star,
i.e., coefficients of the velocity potential of the 1PN order $p$, $q$,
$r$, and $s$. On the other hand, (\ref{Lagdisp6}) --
(\ref{Lagdisp8}) are associated with $e$, $f$, $g$, and $h$.

In order to obtain the orbital angular velocity of 1PN order, we
must calculate the second and third terms on the right-hand side of
Eq. (\ref{preomega}), which come from the displacement of the fluid
element by the 1PN correction. The second term can be evaluated as
follows: First, the contribution from star 1 is zero because
\beqa
  \d \int d^3 x \r {\p U^{1 \ra 1} \over \p x_1} &=&-\d \int d^3 x \r
  ({\bf x}) \int d^3 x' \r ({\bf x}') {x_1 -x'_1 \over | {\bf x} -{\bf
      x}' |^3} \nonumber \\
  &=&-\int d^3 x \r ({\bf x}) \d \int d^3 x' \r ({\bf x}') {x_1 -x'_1
    \over | {\bf x} -{\bf x}' |^3} \nonumber \\
  &&+\int d^3 x' \r ({\bf x}') \d \int d^3 x \r ({\bf x}) {x'_1 -x_1
    \over | {\bf x} -{\bf x}' |^3} \nonumber \\
  &=&0.
\eeqa
The contribution from star 2 can be separated into two parts as
\beqa
  \d \int d^3 x \r {\p U^{2 \ra 1} \over \p x_1} =\int d^3 x \r
  \sum_{i=1}^3 \xi_i {\p^2 U^{2 \ra 1} \over \p x_1 \p x_i} +\int d^3 x
  \r {\p \d U^{2 \ra 1} \over \p x_1}. \label{deltaforce}
\eeqa
The first term on the right-hand side of Eq. (\ref{deltaforce}) denotes
the force to which the displaced element of star 1 is subject from the
nondisplaced potential of star 2. On the other hand, the second term
denotes the force to which the nondisplaced element of star 1 is subject
from the displaced potential of star 2. For the Lagrangian displacement
vectors given in Eqs. (\ref{Lagdisp1}) -- (\ref{Lagdisp8}), the first
and the second terms on the right-hand side of Eq.  (\ref{deltaforce})
are equal, and we can calculate as
\beqa
  \d \int d^3 x \r {\p U \over \p x_1} &=&2 \int d^3 x \r\sum_{i=1}^3
  \xi_i {\p^2 U^{2 \ra 1} \over \p x_1 \p x_i} \nonumber \\
  &=&2\int d^3 x \r {\p \d U^{2 \ra 1} \over \p x_1} \nonumber \\
  &=&2\Bigl[ -{M \over R^4} \Bigl\{ 3S_{11} (2I_{11} +I_{33}) +3S_{12}
  (I_{33} -I_{22}) +S_{31} (2I_{1111} +3I_{1122}) \nonumber \\
  & &\hspace{40pt}+S_{32} (3I_{2233}
  -I_{2222}) -S_{33} (I_{3333} +6I_{1133}) \Bigr\} \nonumber \\
  & &\hspace{15pt}+S_0 M \Bigl( {M \over R^3} +{18\bI_{11} \over R^5}
  \Bigr) \nonumber \\
  & &\hspace{15pt}+S_{21} {M \over
    R^3} \Bigl( I_{11} +{9\bI_{1111} \over R^2} +{9\bI_{11} I_{11} \over
    M R^2} +{12 I_{1133} \over R^2} \Bigr) \nonumber \\
  & &\hspace{15pt}+S_{22} {12M \over R^5} (I_{1133} -I_{1122}) \Bigr].
\eeqa

Also, the third term on the right-hand side of Eq. (\ref{preomega}) is
written as
\beqa
  \O_{\rm DR}^2 F_a (2-F_a) \int d^3 x \r \xi_1
  ={1 \over 2} \O_{\rm DR}^2 F_a (2-F_a) (MS_0 +S_{21} I_{11}).
\eeqa

\subsection{The post-Newtonian orbital angular velocity}

The angular velocity at 1PN order is calculated up to $O(R^{-6})$
from Eq. (\ref{preomega}) as
\beqa
  \d \O^2 &=&{4 \over R^5} \Bigl[ 3S_{11} (2I_{11} +I_{33}) +3S_{12}
  (I_{33} -I_{22}) +S_{31} (2I_{1111} +3I_{1122}) \nonumber \\
  & &\hspace{20pt}+ S_{32} (3I_{2233}
  -I_{2222}) -S_{33} (I_{3333} +6I_{1133}) \Bigr] \nonumber \\
  &&-{2S_0 \over R} \Bigl[ {2M \over R^3} +{36 \bI_{11}
    \over R^5} +{1 \over 2} \O_{\rm DR}^2 F_a (2-F_a) \Bigr] \nonumber
  \\
  &&-{2S_{21} \over R} \Bigl[ {2I_{11}
    \over R^3} +{18\bI_{11} \over 5R^5} a_1^2 +{18\bI_{1111} \over R^5}
  +{24I_{1133} \over R^5} +{1 \over 10} \O_{\rm DR}^2 F_a (2-F_a) a_1^2
  \Bigr] \nonumber \\
  & &-{48 \over R^6} S_{22} (I_{1133} -I_{1122}) \nonumber \\
  &&+2\pi \r \biggl[ \Bigl( {\a_0 \over \pi \r} -{4 \over 5} A_0 +{2
    \over 5} a_1^2 A_1 \Bigr) \Bigl( {M \over R^3} +{9\bI_{11} \over
    2R^5} \Bigr) +{2 \a_1 \over 5R} a_1^2 A_1 \nonumber \\
  &&\hspace{35pt}+{9\bI_{11} \over 2R^5} \Bigl( {P_0 \over \pi \r^2}
  +{\a_0 \over \pi \r} -A_0 \Bigr)
  -\sum_l {\eta_l \over \pi \r} \Bigl( {I_{ll} 
    \over R^3} +{9\bI_{ll11} \over 2R^5} \Bigr) \nonumber \\
  &&\hspace{35pt}-{9\bI_{11} \over 2\pi \r 
    M R^5} \sum_l \eta_l I_{ll} +{1 \over 5} \O_{\rm DR}^2 F_a (2a_2^2
  A_2 +a_3^2 A_3) \nonumber \\
  &&\hspace{35pt}+{3M \over 70R^5} \Bigl\{ \Bigl( -{7P_0 \over \pi
    \r^2} +A_0 +6a_1^2 A_1 \Bigr) (2a_1^2 -a_2^2 -a_3^2) \nonumber \\
  & &\hspace{75pt}+2(2a_1^4 A_1
  -a_2^4 A_2 -a_3^4 A_3) \Bigr\} \biggr] \nonumber \\
  &&-{2M^2 \over R^4} -{2M \over
    R^6} (8I_{11} -3I_{22} -3I_{33}) +{4\a_1 \over R^4} I_{11} \nonumber 
  \\
  &&-{1 \over 8} \O_{\rm DR}^4 R^2 F_a \Bigl\{ 1+{4 \over 5R^2} (3a_1^2
  +a_2^2) F_a^2 \Bigr\} \nonumber \\
  &&-{1 \over 2R} \O_{\rm DR}^2 \Bigl\{ (13F_a -7) M \nonumber \\
  & &\hspace{30pt}-{1
    \over R^2} \Bigl( 2(20F_a^2 -41F_a +21) I_{11} +3(4F_a^2 +5F_a -6)
  I_{22} \nonumber \\
  & &\hspace{70pt}+(13F_a -21) I_{33} \Bigr) \Bigr\} \nonumber \\
  &&+(F_a -1) {2\O_{\rm DR} \over R} \Bigl( e +{3
    \over 5} a_1^2 f +{3 \over 5} a_2^2 g +{1 \over 5}  a_3^2 h \Bigr)
  +{4a_2^2 \over 5R} (F_a +1) \O_{\rm DR} f \nonumber \\
  &&+O(R^{-7}), \\
  &=&{4 \over R^5} \Bigl[ 3S_{11} (2I_{11} +I_{33}) +3S_{12}
  (I_{33} -I_{22}) +S_{31} (2I_{1111} +3I_{1122}) \nonumber \\
  & &\hspace{20pt}+ S_{32} (3I_{2233}
  -I_{2222}) -S_{33} (I_{3333} +6I_{1133}) \Bigr] \nonumber \\
  &&-{2S_0 \over R} \Bigl[ {2M \over R^3} +{36 \bI_{11}
    \over R^5} +{1 \over 2} \O_{\rm DR}^2 F_a (2-F_a) \Bigr] \nonumber
  \\
  &&-{2S_{21} \over R} \Bigl[ {2I_{11}
    \over R^3} +{18\bI_{11} \over 5R^5} a_1^2 +{18\bI_{1111} \over R^5}
  +{24I_{1133} \over R^5} +{1 \over 10} \O_{\rm DR}^2 F_a (2-F_a) a_1^2
  \Bigr] \nonumber \\
  & &-{48 \over R^6} S_{22} (I_{1133} -I_{1122}) \nonumber \\
  &&+(F_a -1) {2\O_{\rm DR} \over R} \Bigl( e +{3
    \over 5} a_1^2 f +{3 \over 5} a_2^2 g +{1 \over 5}  a_3^2 h \Bigr)
  +{4a_2^2 \over 5R} (F_a +1) \O_{\rm DR} f \nonumber \\
  &&+{4M \pi \r \over 5R^3} (4+F_a) A_0 +{9M^2 \over 2R^4} (2-3F_a) +{12
    \pi \r \over 5R^5} (11 +3F_a) A_0 \bI_{11} \nonumber \\
  &&+{M \over R^6} \Bigl[ 2(-3F_a^3 +12F_a^2 -77F_a +54) I_{11}
  +(-10F_a^3 +63F_a -43) I_{22} \nonumber \\
  & &\hspace{30pt}+(55F_a -46) I_{33} \Bigr] +O(R^{-7}).
\eeqa

\section{Boundary conditions}\label{boundary}

In this section, we derive equations to determine the coefficients of
the 1PN velocity potential and the deformation of the binary system,
$p$, $q$, $r$, $s$, $e$, $f$, $g$, $h$, and $S_{ij}$, from the boundary
conditions on the stellar surface. The stellar surfaces of the
Darwin-Riemann ellipsoid and its deformed figure are expressed as
$S_{\rm DR} (x) =0$ and $S (x) =0$, respectively, where
\beqa
  S_{\rm DR} (x) &=& \sum_l {x_l^2 \over a_l^2} -1, \\
  S (x) &=& \sum_l {x_l^2 \over a_l^2} -1 -\sum_j \xi_j {\p S_{\rm DR}
  \over \p x_j}, \\
  &=&S_{\rm DR} (x) -{2 \over c^2} \Bigl[ S_{11} \Bigl(
  {x_1^2 \over a_1^2} -{x_3^2 \over a_3^2} \Bigr) +S_{12} \Bigl( {x_2^2
  \over a_2^2} -{x_3^2 \over a_3^2} \Bigr) +S_{31} \Bigl( {x_1^4 \over
  3a_1^2} -{x_1^2 x_2^2 \over a_2^2} \Bigr) \nonumber \\
  & &\hspace{70pt}+S_{32} \Bigl( {x_2^4 \over
  3a_2^2} -{x_2^2 x_3^2 \over a_3^2} \Bigr) +S_{33} \Bigl( {x_3^4 \over
  3a_3^2} -{x_3^2 x_1^2 \over a_1^2} \Bigr) \nonumber \\
  & &\hspace{70pt}+{1 \over 2} S_0 {x_1 \over
  a_1^2} +S_{21} \Bigl( {x_1^3 \over 2a_1^2} -{x_1 x_3^2 \over a_3^2}
  \Bigr) \nonumber \\
  & &\hspace{70pt}+S_{22} \Bigl( {x_1 x_2^2 \over a_2^2} -{x_1 x_3^2
  \over a_3^2} \Bigr) \Bigr].
\eeqa
The boundary conditions for the continuity equation and integrated Euler 
equation are, respectively,
\beqa
  &&(A) ~~C_i {\p S \over \p x_i} =0~~~~~{\rm on}~~S=0,
  \label{conditionA} \\
  &&(B) ~~{P \over \r} =0~~~~~{\rm on}~~S=0. \label{conditionB}
\eeqa
The condition $(A)$ comes from the constraint that the normal component
of the velocity on the surface vanishes. The condition $(B)$ determines
the stellar surface. Equations (\ref{conditionA}) and (\ref{conditionB}) 
must hold to order $c^{-2}$.

\subsection{Condition (A)}

Substituting $C_i$ given by Eq. (\ref{1PNintvelo}) into
Eq. (\ref{conditionA}), we obtain the equation at 1PN order
\beqa
  0&=&{2\O x_1 x_2 \over c^2} \bigl( \s_0 +\s_1 x_1^2 +\s_2 x^2 +\s_3
  x_3^2 \bigr) \nonumber \\
  &&+{2\O x_2 \over c^2} \bigl( \l_0 +\l_1 x_1^2 +\l_2 x_2^2
  +\l_3 x_3^2 \bigr), \label{condeqa}
\eeqa
where $\s_k$ $(k=0,1,2,3)$ are functions of $a_k$, $p$, $q$, $r$, $s$,
and $S_{ij}$, and also $\l_k$ $(k=0,1,2,3)$ are functions of $a_k$, $e$,
$f$, $g$, $h$, and $S_{ij}$. Equation (\ref{condeqa}) must be satisfied on
$S_{\rm DR}$. Then, we obtain the following six equations (three
equations for $p$, $q$, $r$, $s$, and $S_{ij}$, and three equations for
$e$, $f$, $g$, $h$, and $S_{ij}$):
\beqa
  \Sigma_1^{(A)} &=&{a_1^2 +a_2^2 \over a_1^4 a_2^2} p +{a_1^2 +3a_2^2 \over 
    a_1^2 a_2^2} q +{\O \over a_1^2 +a_2^2} \Bigl[ {4 \over a_1^2}
  (-S_{11} +S_{12}) -{20 \over 3} S_{31} \Bigr] \nonumber \\
  & &-3\pi \r \O (A_0 -a_1^2
  A_1) {a_1^2 -a_2^2 \over a_1^4 a_2^2}
  -{F_a^2 \O^3 \over 2} \Bigl( {a_1^2 -a_2^2 \over a_1^2 a_2^2}
  \Bigr) \nonumber \\
  &&+{\pi \r F_a \O \over 2a_1^2} \Bigl[ \Bigl( {7a_1^2 \over a_2^2}
  -1 \Bigr) B_{11} +\Bigl( {7a_2^2 \over a_1^2} -1 \Bigr) B_{12} +4(a_1^2
  +a_2^2) B_{112} \Bigr] \nonumber \\
  & &-{R^2 \O^3 \over 8} {3a_1^2 +a_2^2 \over a_1^4 a_2^2} F_a +{M\O \over 
    4R} \Bigl( {a_1^2 +11a_2^2 \over a_1^4 a_2^2} \Bigr) \nonumber \\
  & &+{\O \over 2R^3
    a_1^4} \Bigl( {9M \over 2} a_1^2 +2F_a I_{11} +10 F_a I_{22} +{3
    \over 2} I_{22} +{27 \over 4} \bI_{11} \Bigr) \nonumber \\
  & &-{\O \over 2R^3 a_1^2
    a_2^2} \Bigl( -{M \over 2} a_1^2 +14F_a I_{11} -2F_a I_{22} -{3
    \over 2} I_{22} -{81 \over 4} \bI_{11} \Bigr) =0, \label{cond3} \\
  \Sigma_2^{(A)} &=&{a_1^2 +a_2^2 \over a_1^2 a_2^4} p +{3a_1^2 +a_2^2 \over
    a_1^2 a_2^2} r +{\O \over a_1^2 +a_2^2} \Bigl[ {4 \over a_2^2}
  (-S_{11} +S_{12}) +{4a_1^2 \over a_2^2} S_{31} +{8 \over 3} S_{32}
    \Bigr] \nonumber \\
  & &-3\pi \r \O (A_0 -a_2^2 A_2) {a_1^2 -a_2^2 \over a_1^2 a_2^4}
    -{F_a^2 \O^3 \over 2} \Bigl( {a_1^2 -a_2^2 \over a_1^2 a_2^2}
  \Bigr) \nonumber \\
  & &+{\pi \r F_a \O \over 2a_2^2} \Bigl[ \Bigl( {7a_1^2 \over a_2^2}
  -1 \Bigr) B_{12} +\Bigl( {7a_2^2 \over a_1^2} -1 \Bigr) B_{22} +4(a_1^2
  +a_2^2) B_{122} \Bigr] \nonumber \\
  & &-{R^2 \O^3 \over 8} {3a_1^2 +a_2^2 \over a_1^2 a_2^4} F_a +{M\O \over
    4R} \Bigl( {a_1^2 +11a_2^2 \over a_1^2 a_2^4} \Bigr) \nonumber \\
  & &+{\O \over 2R^3
    a_1^2 a_2^2} \Bigl( -{9M \over 4} a_2^2 +2F_a I_{11} +10F_a I_{22}
    +{3 \over 2} I_{22} +{27 \over 4} \bI_{11} \Bigr) \nonumber \\
  & &-{\O \over 2R^3 a_2^4} \Bigl( {21M \over 4} a_2^2 +14F_a I_{11}
    -2F_a I_{22} -{3 \over 2} I_{22} -{81 \over 4} \bI_{11} \Bigr) =0,
    \label{cond4} \\
  \Sigma_3^{(A)} &=&{a_1^2 +a_2^2 \over a_1^2 a_2^2 a_3^2} p +\Bigl( {1
    \over a_1^2} +{1 \over a_2^2} +{2 \over a_3^2} \Bigr) s +{\O
    \over a_1^2 +a_2^2} \Bigl[ {4 \over a_3^2} (-S_{11} +S_{12})
    -{4a_2^2 \over a_3^2} S_{32} +4S_{33} \Bigr] \nonumber \\
  & &-3\pi \r \O (A_0 -a_3^2
    A_3) {a_1^2 -a_2^2 \over a_1^2 a_2^2 a_3^2} \nonumber \\
  & &+{\pi \r F_a \O \over 2a_3^2} \Bigl[ \Bigl( {7a_1^2 \over a_2^2} -1
    \Bigr) B_{13} +\Bigl( {7a_2^2 \over a_1^2} -1 \Bigr) B_{23} +4(a_1^2
    +a_2^2) B_{123} \Bigr] \nonumber \\
  & &-{R^2 \O^3 \over 8} {3a_1^2 +a_2^2 \over a_1^2 a_2^2 a_3^2} F_a +{M\O
    \over 4R} \Bigl( {a_1^2 +11a_2^2 \over a_1^2 a_2^2 a_3^2} \Bigr)
    \nonumber \\
  & &+{\O
    \over 2R^3 a_1^2 a_3^2} \Bigl( -{9M \over 4} a_3^2 +2F_a I_{11}
    +10F_a I_{22} +{3 \over 2} I_{22} +{27 \over 4} \bI_{11} \Bigr)
    \nonumber \\
  & &-{\O \over 2R^3 a_2^2 a_3^2} \Bigl( -{3M \over 2} a_2^2 +{27M \over
    4} a_3^2 +14F_a I_{11} -2F_a I_{22} -{3 \over 2} I_{22} -{81 \over
    4} \bI_{11} \Bigr) =0, \nonumber \\
  & & \label{cond5} \\
  \L_1^{(A)} &=&{e \over a_1^2 a_2^2} +{a_1^2 +2a_2^2 \over a_1^2 a_2^2}
  f +{\O \over a_1^2 +a_2^2} \Bigl[ -{S_0 \over a_1^2} -3S_{21}
  +4S_{22} \Bigr] \nonumber \\
  & &+{R\pi \r \O \over 4a_1^2 a_2^2} (A_0 -a_1^2 A_1 +3a_2^2
  B_{12}) -{R \over 4} F_a^2 \O^3 \Bigl( {3a_1^2 +2a_2^2 \over a_1^2
  a_2^2} \Bigr) \nonumber \\
  & &-{R^3 \O^3 \over 16a_1^2 a_2^2} -{13M\O \over 4a_1^2 a_2^2}
  \nonumber \\
  & &-{\O \over
  2R^2 a_1^2 a_2^2} \Bigl( {M \over 2} a_1^2 +5M a_2^2 -7F_a I_{11}
  +F_a I_{22} +{1 \over 2} I_{22} +{39 \over 4} \bI_{11} \Bigr)
  \nonumber \\
  & &+O(R^{-4}) =0, \label{cond6} \\
  \L_2^{(A)} &=&{e \over a_2^4} +{3g \over a_2^2} -{\O \over a_1^2
    +a_2^2} \Bigl[ {S_0 \over a_2^2} +{2 a_1^2 \over a_2^2} S_{22}
    \Bigr] +{R \pi \r \O \over 4a_2^4} (A_0 -a_2^2 A_2 +3a_2^2 B_{22})
    \nonumber \\
  & &-{R
    F_a^2 \O^3 \over 4a_2^2} -{R^3 \O^3 \over 16a_2^4} -{13M\O \over
    4a_2^4} \nonumber \\
  & &+{\O \over 2R^2 a_2^4} \Bigl( {11M \over 4} a_2^2 +7F_a I_{11}
    -F_a I_{22} -{1 \over 2} I_{22} -{39 \over 4} \bI_{11} \Bigr)
    \nonumber \\
  & &+O(R^{-4}) =0, \label{cond7} \\
  \L_3^{(A)} &=&{e \over a_2^2 a_3^2} +{2a_2^2 +a_3^2 \over a_2^2 a_3^2}
  h +{\O \over a_1^2 +a_2^2} \Bigl[ -{S_0 \over a_3^2} +{2a_1^2
  \over a_3^2} (S_{21} +S_{22}) \Bigr] \nonumber \\
  & &+{R \pi \r \O \over 4a_2^2 a_3^2}
  (A_0 -a_3^2 A_3 +3a_2^2 B_{23}) -{R^3 \O^3 \over 16a_2^2 a_3^2} -{13M\O
  \over 4a_2^2 a_3^2} \nonumber \\
  & &+{\O \over 4R^2 a_2^2 a_3^2} \Bigl( -M a_2^2 +{13M \over 2} a_3^2
  +14F_a I_{11} -2F_a I_{22} -I_{22} -{39 \bI_{11} \over 2} \Bigr)
  \nonumber \\
  & &+O(R^{-4}) =0. \label{cond8}
\eeqa
We can derive the following two equations which do not contain $S_{11}$,
  $S_{12}$ and $p$:
\beqa
  &&\Sigma_1^{(A)} a_1^2 -\Sigma_2^{(A)} a_2^2 =0, \label{relation1} \\
  &&\Sigma_1^{(A)} a_1^2 -\Sigma_3^{(A)} a_3^2 =0. \label{relation2}
\eeqa
Thus, Eqs. (\ref{relation1}) and (\ref{relation2}) depend only on the
  coefficients $q$, $r$, $s$, $S_{31}$, $S_{32}$ and $S_{33}$, i.e., the 
  coefficients for biquadratic deformation. Also, we can construct two
  equations which do not contain $e$ and $S_0$ as
\beqa
  &&\L_1^{(A)} a_1^2 -\L_2^{(A)} a_2^2 =0, \label{relation3} \\
  &&\L_1^{(A)} a_1^2 -\L_3^{(A)} a_3^2 =0. \label{relation4}
\eeqa
Thus, Eqs. (\ref{relation3}) and (\ref{relation4}) depend only on the
  coefficients $f$, $g$, $h$, $S_{21}$ and $S_{22}$, i.e., the
  coefficients for cubic deformation. The latter feature is important in 
  determining coefficients.

\subsection{Condition (B)}

Substituting $\d U$, which is given in Appendix \ref{deltaUform}, and
the surface equation $S=0$ into Eq. (\ref{collecEuler}), we obtain
\beqa
  \Bigl( {P \over \r} \Bigr)_S &=&-{2P_0 \over \r c^2} \Bigl[ S_{11}
  \Bigl( {x_1^2 \over a_1^2} -{x_3^2 \over a_3^2} \Bigr) +S_{12} \Bigl(
  {x_2^2 \over a_2^2} -{x_3^2 \over a_3^2} \Bigr) +S_{31} \Bigl( {x_1^4
    \over 3a_1^2} -{x_1^2 x_2^2 \over a_2^2} \Bigr) \nonumber \\
  & &\hspace{30pt}+S_{32} \Bigl(
  {x_2^4 \over 3a_2^2} -{x_2^2 x_3^2 \over a_3^2} \Bigr)
  +S_{33} \Bigl( {x_3^4 \over 3a_3^2} -{x_3^2 x_1^2
    \over a_1^2} \Bigr) +S_0 {x_1 \over 2a_1^2} \nonumber \\
  & &\hspace{30pt}+S_{21} \Bigl( {x_1^3
    \over 2a_1^2} -{x_1 x_3^2 \over a_3^2} \Bigr) +S_{22} \Bigl( {x_1
    x_2^2 \over a_2^2} -{x_1 x_3^2 \over a_3^2} \Bigr) \Bigr] \nonumber
  \\
  & &+{1 \over c^2} \Bigl[ S_{11} (D_{3,3} -D_{1,1}) +S_{12} (D_{3,3}
  -D_{2,2}) +S_{31} \Bigl( D_{112,2} -{1 \over 3} D_{111,1} \Bigr)
  \nonumber \\
  & &\hspace{25pt}+S_{32} \Bigl( D_{223,3} -{1 \over 3} D_{222,2} \Bigr)
  +S_{33} \Bigl( D_{133,1} -{1 \over 3} D_{333,3} \Bigr) 
  -{1 \over 2} S_0 U_{,1}^{1 \ra 1} \nonumber \\
  & &\hspace{25pt}+S_{21} \Bigl( D_{13,3} -{1 \over 2} 
  D_{11,1} \Bigr) +S_{22} (D_{13,3} -D_{12,2}) \Bigr] \nonumber \\
  & &+{1 \over c^2} \Bigl[ S_{11} (D_{3,3}^{2 \ra 1} -D_{1,1}^{2 \ra 1})
  +S_{12} (D_{3,3}^{2 \ra 1} -D_{2,2}^{2 \ra 1}) \nonumber \\
  & &\hspace{25pt}+S_{31} \Bigl
  ( D_{112,2}^{2 \ra 1} -{1 \over 3} D_{111,1}^{2 \ra 1} \Bigr) +S_{32}
  \Bigl( D_{223,3}^{2 \ra 1} -{1 \over 3} D_{222,2}^{2 \ra 1} \Bigr)
  \nonumber \\
  & &\hspace{25pt}+S_{33} \Bigl( D_{133,1}^{2 \ra 1} -{1 \over 3}
  D_{333,3}^{2 \ra 1} \Bigr) +{1 \over 2} S_0 U_{,1}^{2 \ra 1} \nonumber \\
  & &\hspace{25pt}-S_{21}
  \Bigl( D_{13,3}^{2 \ra 1} -{1 \over 2} D_{11,1}^{2 \ra 1} \Bigr)
  -S_{22} (D_{13,3}^{2 \ra 1} -D_{12,2}^{2 \ra 1}) \Bigr] \nonumber \\
  & &+{1 \over 2c^2} \d \O^2 \Bigl[ F_a (2-F_a) x_1^2 -F_a (2+F_a) x_2^2 
  +Rx_1 +{R^2 \over 4} \Bigr] \nonumber \\
  & &-{1 \over c^2} \Bigl[ \g_0 +\sum_l \g_l x_l^2 +\sum_{l \le m}
  \g_{lm} x_l^2 x_m^2 \nonumber \\
  &&\hspace{25pt}+x_1 \bigl( \k_0 +\sum_l \k_l x_l^2 +\sum_{l \le
    m} \k_{lm} x_l^2 x_m^2 \bigr) \Bigr] +{\rm const} \nonumber \\
  &\equiv&{1 \over c^2} \Bigl[ Q_1 x_1^2 +Q_2 x_2^2 +Q_3
  x_3^2 +Q_{11} x_1^4 +Q_{22} x_2^4 +Q_{33} x_3^4
  +Q_{12} x_1^2 x_2^2 \nonumber \\
  & &\hspace{20pt}+Q_{13} x_1^2 x_3^2 +Q_{23} x_2^2
  x_3^2 \nonumber \\
  & &\hspace{20pt}+x_1 \bigl( R_0 +R_1 x_1^2 +R_2 x_2^2
  +R_3 x_3^2 \bigr) +{\rm const} \Bigr],
\eeqa
where
\beqa
  Q_1 &=& -\g_1 +{1 \over 2} F_a (2-F_a) \d \O^2 +S_{11} \Bigl[ -\pi
  \r \Bigl( {2P_0 \over \pi \r^2 a_1^2} -3a_1^2 A_{11} +a_3^2 A_{13}
  \Bigr) +O(R^{-5}) \Bigr] \nonumber \\
  & &+S_{12} \Bigl[ \pi \r (a_2^2 A_{12} -a_3^2 A_{13}) +O(R^{-5})
  \Bigr] \nonumber \\
  & &+S_{31} \Bigl[ \pi \r a_1^2 \Bigl( -a_1^4 A_{111} +a_1^2 a_2^2
  A_{112} +{3 \over 2} a_1^2 B_{111} -{1 \over 2} a_2^2 B_{112} \Bigr)
  +O(R^{-5}) \Bigr] \nonumber \\
  & &+S_{32} \Bigl[ {1 \over 2} \pi \r a_2^2 (a_2^2 B_{122} -a_3^2
  B_{123}) +O(R^{-5}) \Bigr] \nonumber \\
  & &+S_{33} \Bigl[ {1 \over 2} \pi \r a_3^2 (a_3^2 B_{133} -3a_1^2
  B_{113}) +O(R^{-5}) \Bigr] +O(R^{-4}) \times (S_0,~S_{21}), \\
  Q_2 &=& -\g_2 -{1 \over 2} F_a (2+F_a) \d \O^2 +S_{11} \Bigl[ \pi
  \r (a_1^2 A_{12} -a_3^2 A_{23}) +O(R^{-5}) \Bigr] \nonumber \\
  & &+S_{12} \Bigl[ -\pi \r \Bigl( {2P_0 \over \pi \r^2 a_2^2} -3a_2^2
  A_{22} +a_3^2 A_{23} \Bigr) +O(R^{-5}) \Bigr] \nonumber \\
  & &+S_{31} \Bigl[ {1 \over 2} \pi \r a_1^2 (a_1^2 B_{112} -3a_2^2
  B_{122}) +O(R^{-5}) \Bigr] \nonumber \\
  & &+S_{32} \Bigl[ \pi \r a_2^2 \Bigl( -a_2^4 A_{222} +a_2^2 a_3^2
  A_{223} +{3 \over 2} a_2^2 B_{222} -{1 \over 2} a_3^2 B_{223} \Bigr)
  +O(R^{-5}) \Bigr] \nonumber \\
  & &+S_{33} \Bigl[ {1 \over 2} \pi \r a_3^2 (a_3^2 B_{233} -a_1^2
  B_{123}) +O(R^{-5}) \Bigr] +O(R^{-4}) \times (S_0,~S_{21}), \\
  Q_3 &=& -\g_3 +S_{11} \Bigl[ \pi \r \Bigl( {2P_0 \over \pi \r^2
    a_3^2} -3a_3^2 A_{33} +a_1^2 A_{13} \Bigr) +O(R^{-5}) \Bigr]
    \nonumber \\
  & &+S_{12} \Bigl[ \pi \r \Bigl( {2P_0 \over \pi \r^2 a_3^2} -3a_3^2
    A_{33} +a_2^2 A_{23} \Bigr) +O(R^{-5}) \Bigr] \nonumber \\
  & &+S_{31} \Bigl[ {1 \over 2} \pi \r a_1^2 (a_1^2 B_{113} -a_2^2
    B_{123}) +O(R^{-5}) \Bigr] \nonumber \\
  & &+S_{32} \Bigl[ {1 \over 2} \pi \r a_2^2 (a_2^2 B_{223} -3a_3^2
    B_{233}) +O(R^{-5}) \Bigr] \nonumber \\
  & &+S_{33} \Bigl[ \pi \r a_3^2 \Bigl( -a_3^4 A_{333} +a_3^2 a_1^2
    A_{133} +{3 \over 2} a_3^2 B_{333} -{1 \over 2} a_1^2 B_{133} \Bigr) 
    +O(R^{-5}) \Bigr] \nonumber \\
  & &+O(R^{-4}) \times (S_0,~S_{21}), \\
  Q_{11} &=&-\g_{11} +S_{31} \pi \r \Bigl[ -{2P_0 \over 3\pi \r^2
    a_1^2} +{5 \over 3} a_1^6 A_{1111} -a_1^4 a_2^2 A_{1112} -{5 \over
    4} a_1^4 B_{1111} +{1 \over 4} a_1^2 a_2^2 B_{1112} \Bigr] \nonumber 
    \\
  & &+{1 \over 4} S_{32} \pi \r a_2^2 (a_3^2 B_{1123} -a_2^2 B_{1122})
    +{1 \over 4} S_{33} \pi \r a_3^2 (5a_1^2 B_{1113} -a_3^2 B_{1133}),
    \\
  Q_{12} &=& -\g_{12} +S_{31} \pi \r \Bigl[ {2P_0 \over \pi \r^2
    a_2^2} +a_1^6 A_{1112} -3a_1^4 a_2^2 A_{1122} -{3 \over 2} a_1^4
    B_{1112} +{3 \over 2} a_1^2 a_2^2 B_{1122} \Bigr] \nonumber \\
  & &+S_{32} \pi \r a_2^2 \Bigl( a_2^4 A_{1222} -a_2^2 a_3^2 A_{1223}
    -{3 \over 2} a_2^2 B_{1222} +{1 \over 2} a_3^2 B_{1223} \Bigr)
    \nonumber \\
  &&+{1 \over 2} S_{33} \pi \r a_3^2 (3a_1^2 B_{1123} -a_3^2 B_{1233}), \\
  Q_{13} &=& -\g_{13} +S_{31} \pi \r a_1^2 \Bigl( a_1^4 A_{1113}
  -a_1^2 a_2^2 A_{1123} -{3 \over 2} a_1^2 B_{1113} +{1 \over 2} a_2^2
  B_{1123} \Bigr) \nonumber \\
  & &+{1 \over 2} S_{32} \pi \r a_2^2 (3a_3^2 B_{1233}
  -a_2^2 B_{1223}) \nonumber \\
  & &+S_{33} \pi \r \Bigl[ {2P_0 \over \pi \r^2 a_1^2} +a_3^6 A_{1333}
  -3a_3^4 a_1^2 A_{1133} +{3 \over 2} a_1^2 a_3^2 B_{1133} -{3 \over 2}
  a_3^4 B_{1333} \Bigr], \\
  Q_{22} &=& -\g_{22} +{1 \over 4} S_{31} \pi \r a_1^2 (5a_2^2
  B_{1222} -a_1^2 B_{1122}) \nonumber \\
  & &+S_{32} \pi \r \Bigl[ -{2P_0 \over 3\pi \r^2 a_2^2} +{5 \over 3}
  a_2^6 A_{2222} -a_2^4 a_3^2 A_{2223} -{5 \over 4} a_2^4 B_{2222} +{1
  \over 4} a_2^2 a_3^2 B_{2223} \Bigr] \nonumber \\
  & &+{1 \over 4} S_{33} \pi \r a_3^2 (a_1^2 B_{1223} -a_3^2 B_{2233}),
  \\
  Q_{23} &=& -\g_{23} +{1 \over 2} S_{31} \pi \r a_1^2 (3a_2^2
  B_{1223} -a_1^2 B_{1123}) \nonumber \\
  & &+S_{32} \pi \r \Bigl[ {2P_0 \over \pi \r^2 a_3^2} +a_2^6 A_{2223}
  -3a_2^4 a_3^2 A_{2233} +{3 \over 2} a_2^2 a_3^2 B_{2233} -{3 \over 2}
  a_2^4 B_{2223} \Bigr] \nonumber \\
  & &+S_{33} \pi \r a_3^2 \Bigl( a_3^4 A_{2333} -a_3^2 a_1^2 A_{1233}
  +{1 \over 2} a_1^2 B_{1233} -{3 \over 2} a_3^2 B_{2333} \Bigr), \\
  Q_{33} &=& -\g_{33} +{1 \over 4} S_{31} \pi \r a_1^2 (a_2^2 B_{1233} 
  -a_1^2 B_{1133}) +{1 \over 4} S_{32} \pi \r a_2^2 (5a_3^2 B_{2333}
  -a_2^2 B_{2233}) \nonumber \\
  & &+S_{33} \pi \r \Bigl[ -{2P_0 \over 3\pi \r^2 a_3^2} +{5 \over 3}
  a_3^6 A_{3333} -a_3^4 a_1^2 A_{1333} -{5 \over 4} a_3^4 B_{3333} +{1
  \over 4} a_1^2 a_3^2 B_{1333} \Bigr], \nonumber \\
  && \\
  R_0 &=&-\k_0 +{R \over 2} \d \O^2 +S_0 \Bigl[ \pi \r \Bigl( -{P_0
    \over \pi \r^2 a_1^2} +A_1 \Bigr) +{1 \over 2} \O_{\rm DR}^2 \Bigr]
    \nonumber \\
  & &+S_{21} \Bigl[ \pi \r a_1^2 \Bigl( a_3^2 A_{13} -a_1^2
    A_{11} +{1 \over 2} B_{11} \Bigr) +{I_{11} \over R^3} 
  +O(R^{-5}) \Bigr] \nonumber \\
  & &+S_{22} \Bigl[ \pi \r a_1^2 (a_3^2 A_{13} -a_2^2 A_{12}) 
  +O(R^{-5})
    \Bigr] \nonumber \\
  & &+O(R^{-4}) \times (S_{11},~S_{12},~S_{31},~S_{32},~S_{33}), \\
  R_1 &=& -\k_1 +S_{21} \Bigl[ \pi \r \Bigl( -{P_0 \over \pi \r^2 a_1^2}
  -a_1^2 a_3^2 A_{113} +2a_1^4 A_{111} -{1 \over 2} a_1^2 B_{111} \Bigr)
  +O(R^{-5}) \Bigr] \nonumber \\
  & &+S_{22} \pi \r a_1^2 (a_2^2 A_{112} -a_3^2 A_{113}) +O(R^{-5})
  \times S_0, \\
  R_2 &=&-\k_2 +S_{21} \Bigl[ \pi \r a_1^2 \Bigl(
  a_1^2 A_{112} -a_3^2 A_{123} -{1 \over 2} B_{112} \Bigr) 
  +O(R^{-5}) \Bigr] \nonumber \\
  & &+S_{22} \pi \r \Bigl( -{2P_0 \over \pi \r^2 a_2^2} +3a_1^2 a_2^2
  A_{122} -a_1^2 a_3^2 A_{123} \Bigr)
  +O(R^{-5}) \times S_0, \\
  R_3 &=&-\k_3 +S_{21} \Bigl[ \pi \r \Bigl( {2P_0 
    \over \pi \r^2 a_3^2} +a_1^4 A_{113} -3a_1^2 a_3^2 A_{133} -{1 \over 
    2} a_1^2 B_{113} \Bigr) 
  +O(R^{-5}) \Bigr] \nonumber \\
  & &+S_{22} \pi \r \Bigl( {2P_0 \over \pi \r^2 a_3^2} +a_1^2 a_2^2
    A_{123} -3a_1^2 a_3^2 A_{133} \Bigr)
  +O(R^{-5}) \times S_0.
\eeqa
Since $(P/\r)_S$ must vanish on the stellar surface, we have eight
conditions,
\beqa
  &&a_1^4 Q_{11} +a_2^4 Q_{22} -a_1^2 a_2^2 Q_{12} =0, \label{cond9} \\
  &&a_2^4 Q_{22} +a_3^4 Q_{33} -a_2^2 a_3^2 Q_{23} =0, \label{cond10} \\
  &&a_3^4 Q_{33} +a_1^4 Q_{11} -a_3^2 a_1^2 Q_{13} =0, \label{cond11} \\
  &&a_1^4 Q_{11} -a_2^4 Q_{22} +a_1^2 Q_1 -a_2^2 Q_2 =0, \label{cond12} \\
  &&a_3^4 Q_{33} -a_1^4 Q_{11} +a_3^2 Q_3 -a_1^2 Q_1 =0, \label{cond13}
\eeqa
and
\beqa
  &&R_0 +a_3^2 R_3 =0, \label{cond14} \\
  &&a_1^2 R_1 -a_3^2 R_3 =0, \label{cond15} \\
  &&a_2^2 R_2 -a_3^2 R_3 =0. \label{cond16}
\eeqa
We can see that the equations for determining the coefficients of the
1PN velocity potential and the deformation of the binary system are
separated into two combinations. One of them is constructed from Eqs.
(\ref{cond1}), (\ref{cond3}) -- (\ref{cond5}) and (\ref{cond9}) --
(\ref{cond13}), and determines the coefficients $p$, $q$, $r$, $s$,
$S_{11}$, $S_{12}$, $S_{31}$, $S_{32}$, and $S_{33}$. These coefficients
are associated with the star itself, and the triplane symmetric part of
the Lagrangian displacement vectors. The other is the combination
composed of Eqs. (\ref{cond2}), (\ref{cond6}) -- (\ref{cond8}) and
(\ref{cond14}) -- (\ref{cond16}). These equations determine the
coefficients $e$, $f$, $g$, $h$, $S_0$, $S_{21}$, and $S_{22}$. These
coefficients are associated with the binary motion, and the
$\pi$-rotation symmetric part of the Lagrangian displacement vectors
around the $x_3$-axis.

In the case of an isolated star,\cite{TAS} we have two identical
equations for Eqs. (\ref{cond9}) -- (\ref{cond13}) when we substitute
$\d \O^2$ into Eqs. (\ref{cond12}) and (\ref{cond13}). However, for the
case of a binary system, the five equations (\ref{cond9}) --
(\ref{cond13}) are all independent up to $O(R^{-5})$ even if we
substitute $\d \O^2$ into Eqs. (\ref{cond12}) and (\ref{cond13}).

Contrastingly, when we substitute $\d \O^2$ into Eq. (\ref{cond14}), it
is found that $S_0$ vanishes. This implies that only six of seven
conditions (\ref{cond2}), (\ref{cond6}) -- (\ref{cond8}) and
(\ref{cond14}) -- (\ref{cond16}) are available. Here, we choose
equations excluding Eq. (\ref{cond14}).

There remains the problem of determining the remaining two coefficients
$e$ and $S_0$ because we have only one equation (\ref{cond6}) (or
(\ref{cond7}) or (\ref{cond8})) for these two variables. There is one
degree of freedom. This fact follows from point made by
Chandrasekhar\cite{chandra67,chandra71} and Bardeen\cite{Bardeen} for an
isolated star that there is no unique definition of the 1PN solution as
the counterpart of a Newtonian binary solution.

Since the condition for connecting a 1PN solution to a Newtonian
solution can be arbitrarily chosen, in this paper, we simply give the
condition as
\beqa
  S_0 =0, \label{cond17}
\eeqa
i.e., we fix the 1PN correction to the orbital separation of the binary
system. Then, we have two equations, (\ref{cond6}) and (\ref{cond17}),
for solving for $e$ and $S_0$.

Note that when we calculate Eqs. (\ref{cond12}) and (\ref{cond13}), we
neglect terms such as
\beqa
  {\O_{\rm DR} \over R} F_a \times (e,~f,~g,~h) \label{neglect-term}
\eeqa
in order to separate variables into two groups. We take terms up to
$O(R^{-3})$ in Eqs. (\ref{cond12}) and (\ref{cond13}). In this
calculation, we only count the order of $1/R$ and ignore the order of
the {\it implicit} terms, such as $F_a$ which is $O(R^{-3})$. However,
we regard the term (\ref{neglect-term}) as $O(R^{-4})$. Even if we
include this term, the physical values, i.e., the energy, the angular
momentum and the angular velocity, do not change.\footnote{The change of
  the physical values is less than $1\%$, even near the ISCO.} Note
  that when we include the term (\ref{neglect-term}), only the
  coefficients $p$, $S_{11}$ and $S_{12}$ change. This is
  because we can determine $q$, $r$, $s$, $S_{31}$, $S_{32}$, and
  $S_{33}$ independent of $p$, $S_{11}$ and $S_{12}$ from Eqs.
  (\ref{cond1}), (\ref{relation1}), (\ref{relation2}) and (\ref{cond9})
  -- (\ref{cond11}).

\section{The total energy and angular momentum}

We take the total energy up to $O(R^{-3})$ at Newtonian order and up
to $O(R^{-4})$ at 1PN order. Also the total angular momentum up to
$O(1)$ at Newtonian order and up to $O(R^{-1})$ at 1PN order,
except for $\O_{\rm DR}$.

\subsection{Total energy}

The total energy of the binary system is defined as\cite{chandra65}
\beqa
  E &=&2 \int_V d^3 x \r \Bigl[ {v^2 \over 2} -{1 \over 2} U +{1 \over c^2}
  \Bigl( {5 \over 8} v^4 +{5 \over 2} v^2 U +{1 \over 2} \hat{\b}_i v^i
  +{P \over \r} v^2 -{5 \over 2} U^2 \Bigr) +O(c^{-4}) \Bigr], \nonumber 
  \\
  &=& E_{\rm N} +{1 \over c^2} E_{\rm PN} +O(c^{-4}), \label{totenergy}
\eeqa
where
\beqa
  E_{\rm N} &=&M_{\ast} \Bigl[ {1 \over 5} F_a^2 \O_{\rm DR}^2 (a_{1
    \ast}^2 +a_{2 \ast}^2) +{\t{R}^2 a_{1 \ast}^2 \over 4} \O_{\rm DR}^2
  -{4 \over 5} \pi \r A_{0 \ast} -{M_{\ast} \over \t{R} a_{1 \ast}} -{3
    \bI_{11 \ast} \over \t{R}^3 a_{1 \ast}^3} \Bigr] \nonumber \\
  &&+O(R^{-5}), \\
  E_{\rm PN} &=&E_{\rm N \ra PN} +E_{v^4} +E_{v^2 U^{1 \ra 1}} +E_{v^2
  U^{2 \ra 1}} +E_{v^i \b_i^{1 \ra 1}} +E_{v^i \b_i^{2 \ra 1}} \nonumber 
  \\
  &&+E_{v^2
  P/\r} +E_{(U^{1 \ra 1})^2} +E_{(U^{1 \ra 1})(U^{2 \ra 1})} +E_{(U^{2
  \ra 1})^2}, \\
  E_{\rm N \ra PN}&=&-E_{\rm N} \Bigl[ {1 \over 6} F_a^2 \O_{\rm DR}^2
  (a_{1 \ast}^2 +a_{2 \ast}^2) +4\pi \r A_{0 \ast} +{65M_{\ast} \over 12 
  R} +{75\bI_{11\ast} \over 4R^3} \Bigr] \nonumber \\
  & &+F_a^2 (I_{11\ast} +I_{22\ast}) \d \O^2 +F_a^2 \O_{\rm DR}^2
  (\d I_{11\ast} +\d I_{22\ast}) +{M_{\ast} R^2 \over 4} \d \O^2 \nonumber \\
  & &+F_a \O_{\rm DR}^2 R \d I_{1\ast} \nonumber \\
  & &+2F_a \O_{\rm DR}
  \Bigl\{ p(I_{11\ast} +I_{22\ast}) +q (3I_{1122\ast} +I_{1111\ast})
  \nonumber \\
  &&\hspace{60pt}+r
  (I_{2222\ast} +3I_{1122\ast}) +s (I_{2233\ast} +I_{1133\ast}) \Bigr\}
  \nonumber \\
  & &+R\O_{\rm DR} (eM_{\ast} +f I_{11\ast} +3g I_{22\ast} +h
  I_{33\ast}) +2 \d W_{\ast} \nonumber \\
  & &-\d \int d^3 x \r U^{2 \ra 1}, \label{enpn} \\
  E_{v^4}
  &=&{5 \over 4}\O_{\rm DR}^4 R^4 \Bigl[ {M_{\ast} \over 16} +{1 \over 2R^2}
  F_a^2 (3I_{11\ast} +I_{22\ast}) +O(R^{-4}) \Bigr], \\
  E_{v^2 U^{1 \ra 1}}
  &=&M_{\ast} \pi \r A_{0\ast} \O_{\rm DR}^2 R^2 \Bigl[ 1+{16 F_a^2
  \over 21R^2} (a_{1\ast}^2 +a_{2\ast}^2) +O(R^{-5}) \Bigr], \\
  E_{v^2 U^{2 \ra 1}}
  &=&5\O_{\rm DR}^2 \Bigl[ {M_{\ast}^2 R \over 4} +{M_{\ast} \over R}
  \Bigl\{ F_a^2 (I_{11\ast} +I_{22\ast}) -F_a I_{11\ast} +{3 \over 4}
  \bI_{11\ast} \Bigr\} \nonumber \\
  & &\hspace{30pt}+O(R^{-3}) \Bigr], \\
  E_{v^i \b_i^{1 \ra 1}}
  &=&-{2 \over 35} M_{\ast} \pi \r \O_{\rm DR}^2 F_a^2 a_{1\ast}^2
  a_{2\ast}^2 \Bigl[ \Bigl( {7a_{1\ast}^2 \over a_{2\ast}^2} -1 \Bigr)
  A_{1\ast} +\Bigl( {7a_{2\ast}^2 \over a_{1\ast}^2} -1 \Bigr) A_{2\ast}
  \nonumber \\
  &&\hspace{100pt}+2(a_{1\ast}^2 +a_{2\ast}^2) A_{12\ast} \Bigr] \nonumber \\
  & &-{R^2 \over 10} M_{\ast} \pi \r \O_{\rm DR}^2 (7A_{0\ast}
  +a_{2\ast}^2 A_{2\ast}), \\
  E_{v^i \b_i^{2 \ra 1}}
  &=&{\O_{\rm DR}^2 \over 2} \Bigl[ {7M_{\ast}^2 R \over 4} +{M_{\ast}
  \over 2R} \Bigl( -14F_a I_{11\ast} +2F_a I_{22\ast} +I_{22\ast} +{21
  \over 2} \bI_{11\ast} \Bigr) \nonumber \\
  &&\hspace{30pt}+O(R^{-3}) \Bigr], \\
  E_{v^2 P/ \r}
  &=&{P_0 \over 5\r} M_{\ast} R^2 \O_{\rm DR}^2 \Bigl[ 1 +{4F_a^2 \over 7R^2}
  (a_{1\ast}^2 +a_{2\ast}^2) \Bigr], \\
  E_{(U^{1 \ra 1})^2}
  &=&-{2 \over 7} (\pi \r)^2 M_{\ast} (11A_{0\ast}^2 +a_{1\ast}^4
  A_{1\ast}^2 +a_{2\ast}^4 A_{2\ast}^2 +a_{3\ast}^4 A_{3\ast}^2), \\
  E_{(U^{1 \ra 1})(U^{2 \ra 1})}
  &=&-{8M_{\ast}^2 \over R} \pi \r A_{0\ast} \Bigl[ 1 +{41\bI_{11\ast}
  \over 14R^2 M_{\ast}} +O(R^{-5}) \Bigr], \\
  E_{(U^{2 \ra 1})^2}
  &=&-5 \Bigl[ {M_{\ast}^3 \over R^2} +{M_{\ast}^2 \over R^4}
  (6\bI_{11\ast} +I_{11\ast}) +O(R^{-6}) \Bigr].
\eeqa
In the above equations, we have used the conserved mass
\beqa
  M_{\ast} &\equiv& \int d^3 x \r \Bigl[ 1+{1 \over c^2} \Bigl( {v^2
  \over 2} +3U \Bigr) \Bigr], \nonumber \\
  &=&M \Bigl[ 1+{1 \over c^2} \Bigl\{ {\O_{\rm DR}^2 \over 2}
  \Bigl( {F_a^2 \over 5} (a_1^2 +a_2^2) +{R^2 \over 4} \Bigr) +{12\pi \r 
    \over 5} A_0 +{3M \over R} +{9\bI_{11} \over R^3} \Bigr\}
  \Bigr], \nonumber \\
  &&
\eeqa
instead of the Newtonian mass $M$, and we have defined $\t{R} \equiv
  R/a_1$. Also we substitute the mean radius of the star defined by the
  conserved mass as
\beqa
  a_{\ast} &\equiv& \Bigl( {M_{\ast} \over 4\pi \r/3} \Bigr)^{1/3}
  \nonumber \\
  &=&a_1 (\a_2 \a_3)^{1/3} \Bigl[ 1+{1 \over c^2} \Bigl\{ {\O_{\rm DR}^2
  \over 6} \Bigl( {F_a^2 \over 5} (a_1^2 +a_2^2) +{R^2 \over 4} \Bigr)
  +{4\pi \r \over 5} A_0 \nonumber \\
  &&\hspace{100pt}+{M \over R} +{3\bI_{11} \over R^3} \Bigr\}
  \Bigr],
\eeqa
and $a_{i \ast}$ defined by $a_{1 \ast} a_{2 \ast} a_{3 \ast}
  =a_{\ast}^3$, $a_{2 \ast}/a_{1 \ast} =a_2/a_1$ and $a_{3 \ast}/a_{1
  \ast} =a_3/a_1$. In Eq. (\ref{enpn}), $\d I_1$, $\d I_{ii}$, and $\d
  W$ denote the perturbed dipole moment, the perturbed quadrupole
  moment, and the perturbed self-gravity potential energy,
  respectively. These terms are given in Chandrasekhar's
  textbook\cite{chandra69} and in our previous paper.\cite{TAS} We
  represent these terms in Appendix \ref{perturbedterm}.

The last term on the right-hand side of Eq. (\ref{enpn}) is calculated
from
\beqa
  \d \int d^3 x \r U^{2 \ra 1} =\int d^3 x \r \d U^{2 \ra 1} +\int d^3 x 
  \r \sum_l \xi_l {\p U^{2 \ra 1} \over \p x_l}, \label{deltaW21}
\eeqa
where the first term on the right-hand side of the above equation denotes
the gravitational potential energy which the nondisplaced element of
star 1 experiences due to the displaced potential of star 2. On the other
hand, the second term denotes the gravitational potential energy which
the displaced element of star 1 experiences due to the nondisplaced potential 
of star 2. For the Lagrangian displacement vectors given in
Eqs. (\ref{Lagdisp1}) -- (\ref{Lagdisp8}), the first and second
terms of Eq. (\ref{deltaW21}) are equal, and calculated as
\beqa
  \d \int d^3 x \r U^{2 \ra 1} &=&2\int d^3 x \r \d U^{2 \ra 1},
  \nonumber \\
  &=&2\int d^3 x \r \sum_l \xi_l {\p U^{2 \ra 1} \over \p x_l},
  \nonumber \\
  &=&{2 \over c^2} \biggl[ S_{11} \Bigl\{ {M \over R^3} (2I_{11}
  +I_{33}) +O(R^{-5}) \Bigr\} \nonumber \\
  &&\hspace{15pt}+S_{12} \Bigl\{ {M \over R^3} (I_{33} -I_{22})
  +O(R^{-5}) \Bigr\} \nonumber \\
  &&\hspace{15pt}+S_{31} \Bigl\{ {M \over R^3} \Bigl( {2 \over 3}
  I_{1111} +I_{1122} \Bigr) +O(R^{-5}) \Bigr\} \nonumber \\
  &&\hspace{15pt}+S_{32} \Bigl\{ {M \over
  R^3} \Bigl( I_{2233} -{1 \over 3} I_{2222} \Bigr) +O(R^{-5}) \Bigr\}
  \nonumber \\
  &&\hspace{15pt}-S_{33} \Bigl\{ {M \over
  R^3} \Bigl( 2I_{1133} +{1 \over 3} I_{3333} \Bigr) +O(R^{-5}) \Bigr\}
  \nonumber \\
  &&\hspace{15pt}-{1 \over 2} S_0  \Bigl\{ {M^2 \over R^2} +{9M \bI_{11}
  \over R^4} +O(R^{-6}) \Bigr\} \nonumber \\
  &&\hspace{15pt}-S_{21} \Bigl\{ {MI_{11} \over 2R^2} +{9I_{11} \over
  4R^4} \bI_{11} +{9M \over 4R^4} \bI_{1111} +{3M \over R^4} I_{1133}
  +O(R^{-6}) \Bigr\} \nonumber \\
  &&\hspace{15pt}-S_{22} \Bigl\{ {3M \over R^4} (I_{1133} -I_{1122})
  +O(R^{-6}) \Bigr\} \biggr].
\eeqa

\subsection{Total angular momentum}

The total angular momentum is written as\cite{chandra65}
\beqa
  J &=&2 \int_V d^3 x \r \Bigl[ v_{\vp} \Bigl\{ 1+{1 \over c^2} \Bigl( v^2 
  +6U +{P \over \r} \Bigr) \Bigr\} +{\hat{\b}_{\vp} \over c^2}
  +O(c^{-4}) \Bigr], \nonumber \\
  &=&J_{\rm N} +{1 \over c^2} J_{\rm PN} +O(c^{-4}), \label{totangmom}
\eeqa
where
\beqa
  v_{\vp} &=& -x_2 v_1 +\Bigl( x_1 +{R \over 2} \Bigr) v_2, \\
  \hat{\b}_{\vp} &=& -x_2 \hat{\b}_1 +\Bigl( x_1 +{R \over 2} \Bigr)
  \hat{\b}_2, \\
  J_{\rm N}
  &=&2\O_{\rm DR} M_{\ast} \Bigl[ {F_a \over 5} (a_{1 \ast}^2 -a_{2
    \ast}^2) +{\t{R}^2 \over 4} a_{1 \ast}^2 \Bigr], \\
  J_{\rm PN} &=&J_{\rm N \ra PN} +J_{v_{\vp} v^2} +J_{v_{\vp} U^{1 \ra 1}} 
  +J_{v_{\vp} U^{2 \ra 1}} +J_{v_{\vp} P/ \r} \nonumber \\
  &&+J_{\b_{\vp}^{1 \ra 1}}+J_{\b_{\vp}^{2 \ra 1}}, \\
  J_{\rm N \ra PN} &=&-J_{\rm N} \Bigl[ {1 \over 6} F_a^2 \O_{\rm DR}^2
  (a_{1 \ast}^2 +a_{2 \ast}^2) +4\pi \r A_{0 \ast} +{65M_{\ast} \over
  12R} +{75\bI_{11\ast} \over 4R^3} \Bigr] \nonumber \\
  & &+2\Bigl[ F_a (I_{11\ast} -I_{22\ast}) \d \O +F_a \O_{\rm DR} (\d
  I_{11\ast} -\d I_{22\ast}) +{M_{\ast}R^2 \over 4} \d \O \nonumber \\
  &&\hspace{15pt}+{R \over 2}(F_a +1) \O_{\rm DR} \d I_{1\ast}
  +p(I_{11\ast} -I_{22\ast}) \nonumber \\
  & &\hspace{15pt}+q(I_{1111\ast} -3I_{1122\ast}) +r(3I_{1122\ast}
  -I_{2222\ast}) +s(I_{1133\ast} -I_{2233\ast}) \nonumber \\
  & &\hspace{15pt}+{R \over 2} (eM_{\ast}
  +f I_{11\ast} +3g I_{22\ast} +hI_{33\ast}) \Bigr], \\
  J_{v_{\vp} v^2}
  &=&{R^4 \over 2} \O_{\rm DR}^3 \Bigl[ {M_{\ast} \over 4} +{F_a \over R^2}
  \Bigl\{ 3(F_a +1) I_{11\ast} +(F_a -1) I_{22\ast} \Bigr\} \nonumber \\
  &&\hspace{35pt}+O(R^{-4}) \Bigr],
  \\
  J_{v_{\vp} U^{1 \ra 1}}
  &=&{12 \over 5} M_{\ast} \pi \r \O_{\rm DR} R^2 A_{0\ast} \Bigl
  [ 1+{16F_a \over 21R^2} (a_{1\ast}^2 -a_{2\ast}^2) +O(R^{-5}) \Bigr], \\
  J_{v_{\vp} U^{2 \ra 1}}
  &=&3 \O_{\rm DR} \Bigl[ M_{\ast}^2 R +{M_{\ast} \over R}
  \Bigl\{ 2(F_a -1) I_{11\ast} -4F_a I_{22\ast} +3\bI_{11\ast} \Bigr\}
  \nonumber \\
  &&\hspace{35pt}+O(R^{-3}) \Bigr], \\
  J_{v_{\vp} P/ \r}
  &=&{P_0 \over 5\r} \O_{\rm DR} M_{\ast} R^2 \Bigl[ 1+{4F_a \over 7R^2}
  (a_{1\ast}^2 -a_{2\ast}^2) \Bigr], \\
  J_{\b_{\vp}^{1 \ra 1}}
  &=&-{32 \over 105} M_{\ast} \pi \r \O_{\rm DR} F_a \Bigl[ A_{0\ast}
  (a_{1\ast}^2 -a_{2\ast}^2) +O(R^{-3}) \Bigr] \nonumber \\
  &&-{R^2 \over 5} M_{\ast} \pi \r \O_{\rm DR} (7A_{0\ast} +a_{2\ast}^2
  A_{2\ast}), \\
  J_{\b_{\vp}^{2 \ra 1}}
  &=&\O_{\rm DR} \Bigl[ {7M_{\ast}^2 R \over 4} +{M_{\ast} \over 2R}
  \Bigl\{ -7(F_a +1) I_{11\ast} +F_a I_{22\ast} +{21\bI_{11 \ast} \over
  2} \Bigr\} \nonumber \\
  &&\hspace{25pt}+O(R^{-3}) \Bigr],
\eeqa
and
\beqa
  \d \O ={\d \O^2 \over 2\O_{\rm DR}}.
\eeqa

\section{Numerical results}

In the following subsections, we calculate the equilibrium sequence of
the irrotational binary system.

\subsection{Normalization}

First of all, we introduce non-dimensional parameters according to
\beqa
  &&\t{p} \equiv{p \over (M_{\ast}^3/a_{\ast}^5)^{1/2}},
  ~~\t{q} \equiv{q \over (M_{\ast}/a_{\ast}^3)^{3/2}},
  ~~\t{r} \equiv{r \over (M_{\ast}/a_{\ast}^3)^{3/2}},
  ~~\t{s} \equiv{s \over (M_{\ast}/a_{\ast}^3)^{3/2}},~~~~~ \\
  &&\t{e} \equiv{e \over (M_{\ast}/a_{\ast})^{3/2}},
  ~~\t{f} \equiv{f \over (M_{\ast}^3/a_{\ast}^7)^{1/2}},
  ~~\t{g} \equiv{g \over (M_{\ast}^3/a_{\ast}^7)^{1/2}},
  ~~\t{h} \equiv{h \over (M_{\ast}^3/a_{\ast}^7)^{1/2}},~~~~~ \\
  &&\t{S}_{11} \equiv{S_{11} \over M_{\ast}/a_{\ast}},
  ~~~\t{S}_{12} \equiv{S_{12} \over M_{\ast}/a_{\ast}}, \nonumber \\
  &&\t{S}_{31} \equiv{S_{31} \over M_{\ast}/a_{\ast}^3},
  ~~~\t{S}_{32} \equiv{S_{32} \over M_{\ast}/a_{\ast}^3},
  ~~~\t{S}_{33} \equiv{S_{33} \over M_{\ast}/a_{\ast}^3}, \\
  &&\t{S}_0 \equiv{S_0 \over M_{\ast}},
  ~~~\t{S}_{21} \equiv{S_{21} \over M_{\ast}/a_{\ast}^2},
  ~~~\t{S}_{22} \equiv{S_{22} \over M_{\ast}/a_{\ast}^2}, \\
  &&\t{\O}_{\rm DR}^2 \equiv{\O_{\rm DR}^2 \over M_{\ast}/a_{\ast}^3},
  ~~~\d \t{\O}^2 \equiv{\d \O^2 \over M_{\ast}^2/a_{\ast}^4},
  ~~~\d \t{\O} \equiv{\d \O \over (M_{\ast}^3/a_{\ast}^5)^{1/2}},
  ~~~\t{P}_0 \equiv{P_0 \over \r M_{\ast}/a_{\ast}}.
\eeqa
Using these parameters with the condition $S_0=0$, we can rewrite
two groups of equations into non-dimensional forms. One of these is
(\ref{cond1}), (\ref{cond3}) -- (\ref{cond5}), (\ref{cond9}) --
(\ref{cond13}), and the other is (\ref{cond2}), (\ref{cond6}) --
(\ref{cond8}), (\ref{cond15}) and (\ref{cond16}). {}From these
fifteen equations, we can determine fifteen variables, $\t{p}$, $\t{q}$, 
$\t{r}$, $\t{s}$, $\t{S}_{11}$, $\t{S}_{12}$, $\t{S}_{31}$, $\t{S}_{32}$ 
and $\t{S}_{33}$, and $\t{e}$, $\t{f}$, $\t{g}$, $\t{h}$, $\t{S}_{21}$
and $\t{S}_{22}$.

After determination of the variables, we can calculate the total energy
and total angular momentum of the binary system from
Eqs. (\ref{totenergy}) and (\ref{totangmom}). In the numerical
calculations, these values must be normalized. The normalized energy and 
angular momentum are written as
\beqa
  \t{E} &\equiv& {E \over M_{\ast}^2/a_{\ast}} =\t{E}_{\rm N} +C_{\rm s}
  \t{E}_{\rm PN}, \\
  \t{J} &\equiv& {J \over (M_{\ast}^3 a_{\ast})^{1/2}} =\t{J}_{\rm N}
  +C_{\rm s} \t{J}_{\rm PN},
\eeqa
where $C_{\rm s}$ is the compactness parameter defined by
\beqa
  C_{\rm s} \equiv {M_{\ast} \over c^2 a_{\ast}}.
\eeqa
We consider the case $C_{\rm s} \ll 1$ because of the 1PN approximation.

At 1PN order, the center of mass of the binary system deviates from
the value defined at Newtonian order.
The center of mass at 1PN order is defined by using the conserved
mass as
\beqa
  x^i_{\ast} \equiv {1 \over M_{\ast}} \int d^3 x \r_{\ast} x^i.
\eeqa
The $x_1$ component is
\beqa
  x^1_{\ast} ={1 \over M_{\ast}} \Bigl[ \d I_1 +{1 \over c^2} \Bigl\{
  \Bigl( {\O_{\rm DR}^2 \over 2} F_a R -{3M \over R^2} \Bigr) I_{11}
  +O(R^{-4}) \Bigr\} \Bigr],
\eeqa
Then, the orbital separation of the 1PN order is written as
\beqa
  R_{\ast} =R \Bigl[ 1+{1 \over c^2} \Bigl\{ {a_1^2 \over 5} \Bigl(
  \O_{\rm DR}^2 F_a -{6M \over R^3} \Bigr) +{1 \over R} \Bigl( S_0 +{1
    \over 5} S_{21} a_1^2 \Bigr) \Bigr\} \Bigr].
\eeqa

Once an equilibrium sequence is obtained, we search for the minimum
point of the total energy. If we find it, we call it the ISCO.

\subsection{Ellipsoidal approximation}

The ellipsoidal approximation, in which the equilibrium configuration is
assumed to have an ellipsoidal figure, is useful when we study features
of stars or binary systems in the 1PN approximation,\cite{Lom,TS,SZ}
because the whole calculation is carried out by setting $S_{ij}=0$, and
hence is greatly simplified. The ellipsoidal approximation gives an
exact solution for a rotating incompressible star in the Newtonian
theory. However, it gives only an approximate solution for the 1PN
case. If the ellipsoidal approximation is truly a robust approximation,
it becomes a useful method for the study of the 1PN effects. This is the
motivation for our investigation of the validity of the ellipsoidal
approximation.

In the ellipsoidal approximation, a solution is obtained by setting
$S_{ij}=0$ in all the equations. After we set $S_{ij}=0$, we can
calculate the velocity potential at 1PN order from Eqs. (\ref{cond1})
and (\ref{cond3}) -- (\ref{cond5}), and also (\ref{cond2}) and
(\ref{cond6}) -- (\ref{cond8}). We note that in the ellipsoidal
approximation, the boundary condition (\ref{conditionB}) is not
satisfied.

\subsection{Results}

The results are shown in Figs. 2 -- 5. In Fig. 2, we represent the total
energy and total angular momentum as functions of the normalized
separation of the binary system $R_{\ast}/a_{\ast}$ and the normalized
orbital angular velocity $\O/\sqrt{M_{\ast}/a_{\ast}^3}$. We can see
from this figure that the minimum point of the total angular momentum is
slightly different from that of the total energy. The deviation between
the location of the minimums of the energy and angular momentum comes
from the fact that we assume $a_{\ast}/R_{\ast}$ is a small parameter
and expand the energy up to $O(R_{\ast}^{-4})$ and the angular momentum
up to $O(R_{\ast}^{-1})$. In any case, we may expect that the ISCO is
located near the two minimums. It is found that even if we increase the
compactness parameter $C_{\rm s}$, the angular velocity at the ISCO in
units of $\sqrt{M_{\ast}/a_{\ast}^3}$ has almost the same value as in
the Newtonian case, while the orbital separation decreases. This feature
of the angular velocity at the ISCO is different from that in the case
of the corotational binary system. In such a case, we have obtained the
result that when we increase the compactness parameter, the angular
velocity at the ISCO increases.\cite{TS} In a previous paper
investigating irrotational and incompressible stars,\cite{TAS} we found
the instability driven by the deformation of 1PN order. We regard such
an instability point as the bifurcation point to a new sequence.
However, in the present study, we do not find such an instability point
throughout the equilibrium sequence of the binary system until the ISCO.

\begin{figure}[ht]
\epsfxsize 9cm 
\begin{center}
\leavevmode
\epsfbox{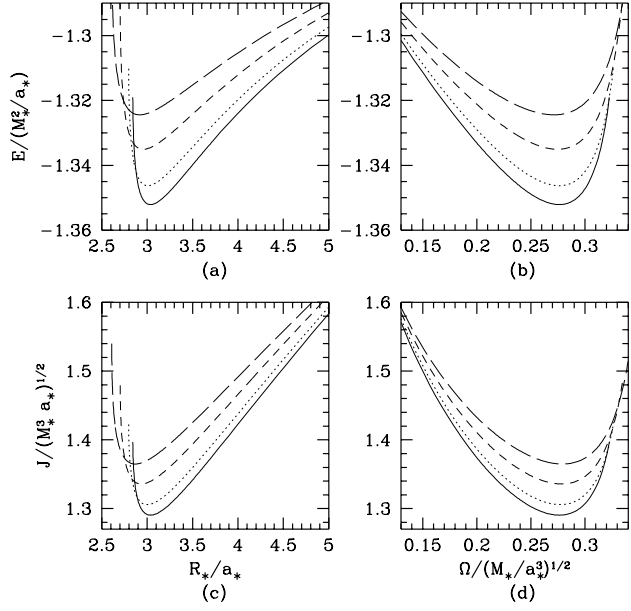}
\end{center}
\caption{The total energy and the total angular momentum of the {\it full}
  calculation as functions of the orbital separation $R_{\ast}/a_{\ast}$
  and the angular velocity $\O/(M_{\ast}/a_{\ast}^3)^{1/2}$. The solid,
  dotted, dashed, and long-dashed lines denote the cases of $C_{\rm
  s}=0$ (Newtonian), $0.01$, $0.03$, and $0.05$, respectively.
}
\end{figure}

\begin{figure}[ht]
\epsfxsize 9cm 
\begin{center}
\leavevmode
\epsfbox{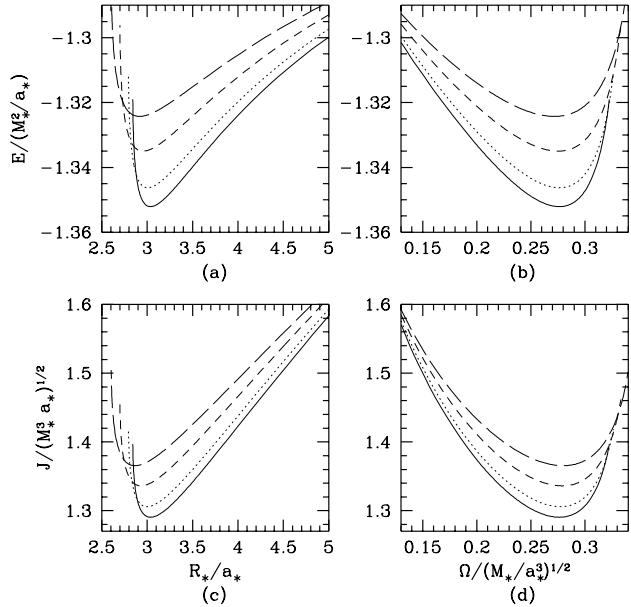}
\end{center}
\caption{The total energy and the total angular momentum of the
  {\it ellipsoidal} approximation as functions of the orbital separation
  $R_{\ast}/a_{\ast}$ and the angular velocity
  $\O/(M_{\ast}/a_{\ast}^3)^{1/2}$. The identifications of the lines are
  the same as in Fig. 2.
}
\end{figure}

In Fig. 3, the total energy and total angular momentum in the
ellipsoidal approximation are shown as functions of the normalized
separation of the binary system and the normalized orbital angular
velocity. It is found that the quantitative features of the energy and
angular momentum are the same as in Fig. 2. Also, we find that values at
1PN order ($\d \O_{\rm PNe}$, $E_{\rm PNe}$ and $J_{\rm PNe}$) differ by
only a few percent outside the ISCO from those of the full calculation,
even if the figure is assumed to be an ellipsoid. This implies that the
ellipsoidal approximation gives good approximate results.

\begin{figure}[ht]
\epsfxsize 9cm 
\begin{center}
\leavevmode
\epsfbox{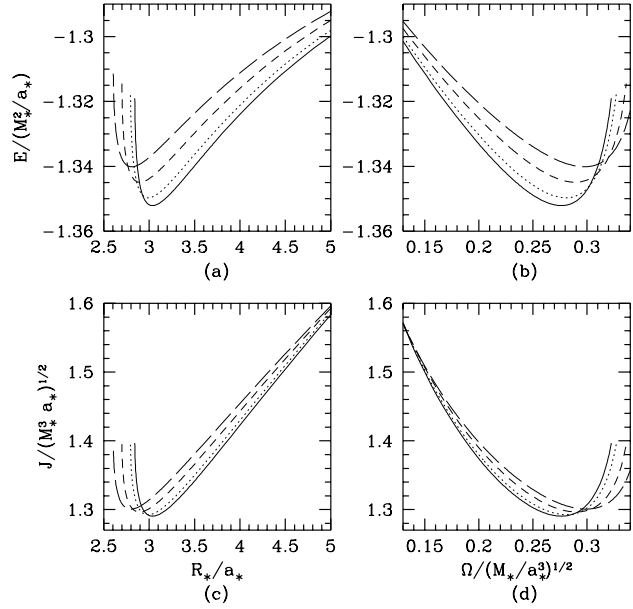}
\end{center}
\caption{The total energy and the total angular momentum of the
  approximation in which we neglect the velocity potential at 1PN
  order as functions of the orbital separation
  $R_{\ast}/a_{\ast}$ and the angular velocity
  $\O/(M_{\ast}/a_{\ast}^3)^{1/2}$. The identifications of the lines are
  the same as in Fig. 2.
}
\end{figure}

\begin{figure}[ht]
\epsfxsize 9cm 
\begin{center}
\leavevmode
\epsfbox{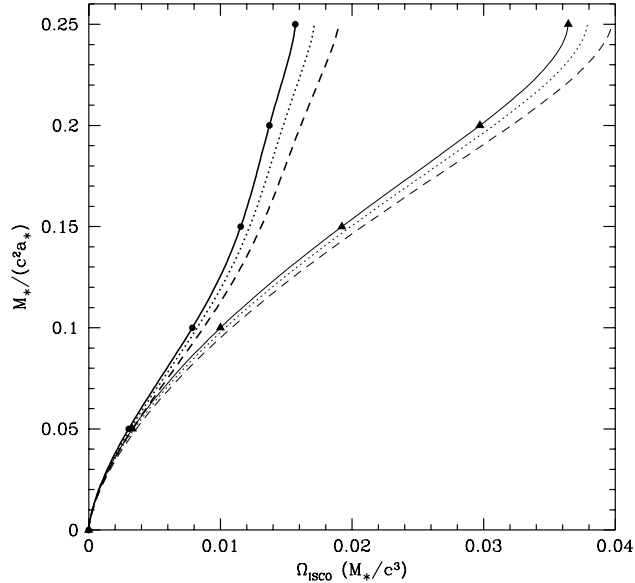}
\end{center}
\caption{The relation between the angular velocity at the ISCO $\O_{\rm
    ISCO} (M_{\ast}/c^3)$ and the compactness parameter $C_{\rm s}
  =M_{\ast}/c^2 a_{\ast}$. The thick and thin lines denote the cases of
  the full calculation and the absence of the 1PN velocity potential,
  respectively. Filled circles and triangles denote the results of our
  calculation. Solid, dotted and dashed lines are the fitting lines for
  $n=0$ (incompressible case), $0.5$ and $0.7$, where $n$ denotes the
  polytropic index.
}
\end{figure}

However, neglect of the velocity potential of 1PN order has an
effect on the quantitative nature of the energy and angular momentum.
We can see from Fig. 4 that the angular velocity at the minimum point of
the energy and angular momentum increases when we increase the
compactness parameter. Figure 4 is obtained by setting the velocity
potential at 1PN order to zero and solving only Eqs.  (\ref{cond3})
-- (\ref{cond5}) and (\ref{cond9}) -- (\ref{cond13}), and also
(\ref{cond6}) -- (\ref{cond8}), (\ref{cond15}) and (\ref{cond16}).

In Fig. 5, we show the relation between the compactness parameter
$C_{\rm s} =M_{\ast}/c^2 a_{\ast}$ and the angular velocity at the ISCO
in units of $c^3/M_{\ast}$. Although the 1PN approximation is not valid,
we take $C_{\rm s}$ up to $0.25$, which we regard as the realistic
compactness parameter of a neutron star, in order to compare our results
with those of Lombardi, Rasio and Shapiro\cite{Lom} (see Fig. 4 in their
paper). In Fig. 5, the thick and thin solid lines denote the cases of
the full calculation and the absence of the 1PN velocity potential.
These lines come from our calculation for an incompressible fluid. The
dotted and dashed lines denote the ``compressible'' cases, whose
polytropic indices are $0.5$ and $0.7$, respectively. We plot these
lines by assuming that the 1PN correction to the angular velocity is the
same value as in the incompressible case if the stars in the binary
system composed of compressible fluid have a polytropic index less than
$1$, i.e., the hard equation of state. Also, we adopt the angular
velocity at the ISCO given by Ury\=u and Eriguchi\cite{UE2} as that of
the irrotational and compressible binary system in the Newtonian theory.

We can see from Fig. 5 that the angular velocity at the ISCO is
overestimated when we neglect the 1PN velocity potential.  Figure 5
agrees qualitatively with that of Lombardi, Rasio and Shapiro.\cite{Lom}
However, our results suggest that the angular velocities given by
Lombardi, Rasio and Shapiro are overestimated because of the absence of
the 1PN velocity potential in their calculation. The reason that the
total force of the system balances with the smaller orbital angular
velocity than that in the case of the absence of the 1PN velocity
potential is as follows. We regard the energy minimum point as the ISCO
in this paper. As the stars in the binary system become closer and
closer to each other, the energy terms with a minus sign, such as the
binding energy term, decrease and those with a plus sign, such as the
rotation energy term, increase. At the ISCO, the increase of the energy
terms with a plus sign is equal to the decrease of those with a minus
sign. When the star possesses internal motion, such motion can produce a
part of the energy with a plus sign. Therefore, if the star with the 1PN
velocity potential has the same ellipsoidal figure as that without the
1PN velocity potential, the energy of the orbital motion, that is to
say, the orbital angular velocity, can decrease.

Finally, we summarize the results in Tables I -- VI. In Table I, the
angular velocity, energy and angular momentum of the Newtonian and 1PN
orders are shown along the sequence of the binary system. In this table,
the symbol $\dagger$ denotes the location of the ISCO of the Newtonian
binary system, i.e., the point at which the total energy has its
minimum.

In Table II, the total angular velocity, total energy, and total angular
momentum at 1PN order are shown. The symbol $\dagger$ denotes the
location of the ISCO. In Table III, we give the coefficients of the
velocity potential and deformation associated with the star itself, and
in Table IV those associated with the binary motion are represented.
Also, we show the results of the ellipsoidal approximation in Tables V
and VI. We can see from Table V that the values of the energy, angular
momentum and angular velocity at 1PN order in the ellipsoidal
approximation deviate by only a few percent from those in the full
calculation (Table I).

\begin{table}
 \caption{Equilibrium sequences of the irrotational Darwin-Riemann
   ellipsoids at Newtonian order and the angular velocity, the
   energy, and the angular momentum at 1PN order along the
   sequences. The symbol $\dagger$ denotes the energy minimum point of the
   Newtonian sequence, i.e., the ISCO defined in this paper.
}
 \begin{center}
  \begin{tabular}{cc|cc|ccc|ccc}
   $R/a_1$&$R_{\ast}/a_{\ast}$&$a_2/a_1$&$a_3/a_1$&
   $\t{\O}_{\rm DR}$&$\t{E}_{\rm N}$&$\t{J}_{\rm N}$&
   $\d \t{\O}$&$\t{E}_{\rm PN}$&$\t{J}_{\rm PN}$ \\ \hline
   6.00&6.071&0.9824&0.9828&~~~9.461(-2)&-1.282&1.743&
   0.1249&~~~2.944(-2)&-0.7523 \\
   5.50&5.584&0.9771&0.9777&0.1073&-1.289&1.673&
   0.1401&~~~5.964(-2)&-0.4832 \\
   5.00&5.103&0.9693&0.9705&0.1229&-1.298&1.600&
   0.1584&~~~9.933(-2)&-0.1947 \\
   4.50&4.628&0.9576&0.9597&0.1424&-1.308&1.525&
   0.1806&0.1523&~0.1157 \\
   4.00&4.165&0.9391&0.9430&0.1671&-1.319&1.450&
   0.2079&0.2239&~0.4498 \\
   3.50&3.722&0.9078&0.9160&0.1986&-1.333&1.377&
   0.2412&0.3212&~0.8081 \\
   3.00&3.317&0.8514&0.8692&0.2382&-1.346&1.315&
   0.2802&0.4511&1.190~ \\
   $\dagger$2.582&3.037&0.7671&0.8013&0.2763&-1.352&1.291&
   0.3147&0.5879&1.548~ \\
   2.50&2.992&0.7446&0.7831&0.2839&-1.352&1.292&
   0.3215&0.6225&1.641~ \\
   2.00&2.842&0.5586&0.6238&0.3226&-1.319&1.397&
   0.3984&0.9807&2.856~ \\
  \end{tabular}
 \end{center}
\end{table}%

\begin{table}
 \caption{Equilibrium sequences of the irrotational Darwin-Riemann
   ellipsoids at 1PN order with the compactness parameters
   $M_{\ast}/c^2 a_{\ast}=0.01$, $0.03$, and
   $0.05$. The symbol $\dagger$ denotes the point of the ISCO.
}
 \begin{center}
  \begin{tabular}{ccccccc}
   $R/a_1$&$R_{\ast}/a_{\ast}$&$a_2/a_1$&$a_3/a_1$&$\t{\O}$&
   $\t{E}$&$\t{J}$ \\ \hline \hline
   \multicolumn{7}{c}{$M_{\ast}/c^2a_{\ast}=0.01$} \\ \hline
   6.00&5.987&0.9824&0.9828&~~~9.585(-2)&-1.282&1.736 \\
   5.50&5.506&0.9771&0.9777&0.1087&-1.289&1.668 \\
   5.00&5.030&0.9693&0.9705&0.1244&-1.297&1.598 \\
   4.50&4.561&0.9576&0.9597&0.1442&-1.306&1.526 \\
   4.00&4.104&0.9391&0.9430&0.1691&-1.317&1.454 \\
   3.50&3.666&0.9078&0.9160&0.2010&-1.329&1.385 \\
   3.00&3.265&0.8514&0.8692&0.2410&-1.341&1.327 \\
   $\dagger$2.618&3.008&0.7762&0.8086&0.2761&-1.346&1.306 \\
   2.50&2.944&0.7446&0.7831&0.2871&-1.346&1.308 \\
   2.00&2.796&0.5586&0.6238&0.3266&-1.309&1.426 \\ \hline \hline
   \multicolumn{7}{c}{$M_{\ast}/c^2a_{\ast}=0.03$} \\ \hline
   6.00&5.819&0.9824&0.9828&~~~9.835(-2)&-1.281&1.721 \\
   5.50&5.350&0.9771&0.9777&0.1115&-1.288&1.658 \\
   5.00&4.885&0.9693&0.9705&0.1276&-1.295&1.594 \\
   4.50&4.428&0.9576&0.9597&0.1478&-1.303&1.529 \\
   4.00&3.981&0.9391&0.9430&0.1733&-1.313&1.463 \\
   3.50&3.553&0.9078&0.9160&0.2058&-1.323&1.401 \\
   3.00&3.161&0.8514&0.8692&0.2466&-1.332&1.351 \\
   $\dagger$2.698&2.958&0.7951&0.8238&0.2747&-1.335&1.336 \\
   2.50&2.848&0.7446&0.7831&0.2936&-1.333&1.341 \\
   2.00&2.703&0.5586&0.6238&0.3346&-1.290&1.483 \\ \hline \hline
   \multicolumn{7}{c}{$M_{\ast}/c^2a_{\ast}=0.05$} \\ \hline
   6.00&5.651&0.9824&0.9828&0.1009&-1.281&1.706 \\
   5.50&5.193&0.9771&0.9777&0.1143&-1.286&1.648 \\
   5.00&4.740&0.9693&0.9705&0.1308&-1.293&1.590 \\
   4.50&4.294&0.9576&0.9597&0.1514&-1.300&1.531 \\
   4.00&3.858&0.9391&0.9430&0.1775&-1.308&1.472 \\
   3.50&3.440&0.9078&0.9160&0.2106&-1.316&1.417 \\
   3.00&3.057&0.8514&0.8692&0.2522&-1.323&1.374 \\
   $\dagger$2.795&2.919&0.8156&0.8403&0.2715&-1.324&1.366 \\
   2.50&2.751&0.7446&0.7831&0.3000&-1.321&1.373 \\
   2.00&2.610&0.5586&0.6238&0.3425&-1.270&1.540 \\
  \end{tabular}
 \end{center}
\end{table}%

\begin{table}
 \caption{Coefficients of the 1PN velocity potential and the Lagrangian
   displacement vectors associated with the star itself shown along the
   equilibrium sequence of the irrotational Darwin-Riemann ellipsoid.
}
 \begin{center}
  {\tabcolsep=2pt
  \begin{tabular}{c|cccc|ccccc}
   $R/a_1$&$\t{p}$&$\t{q}$&$\t{r}$&$\t{s}$&
   $\t{S}_{11}$&$\t{S}_{12}$&$\t{S}_{31}$&$\t{S}_{32}$&$\t{S}_{33}$ \\
   \hline
   6.00&-1.634(-2)&-6.281(-4)&~5.725(-5)&~4.333(-5)&
   ~3.205(-3)&-1.997(-2)&-1.950(-4)&-1.851(-5)&6.821(-5) \\
   5.50&-1.937(-2)&-9.253(-4)&~8.352(-5)&~5.711(-5)&
   ~8.525(-4)&-1.978(-2)&-3.235(-4)&-3.191(-5)&1.152(-4) \\
   5.00&-2.296(-2)&-1.412(-3)&~1.256(-4)&~7.251(-5)&
   -2.921(-3)&-1.884(-2)&-5.599(-4)&-5.826(-5)&2.046(-4) \\
   4.50&-2.686(-2)&-2.248(-3)&~1.948(-4)&~8.096(-5)&
   -8.970(-3)&-1.665(-2)&-1.016(-3)&-1.141(-4)&3.862(-4) \\
   4.00&-2.990(-2)&-3.760(-3)&~3.104(-4)&~4.680(-5)&
   -1.855(-2)&-1.258(-2)&-1.943(-3)&-2.436(-4)&7.863(-4) \\
   3.50&-2.752(-2)&-6.660(-3)&~4.937(-4)&-1.711(-4)&
   -3.264(-2)&-6.665(-3)&-3.896(-3)&-5.781(-4)&1.760(-3) \\
   3.00&-1.683(-3)&-1.249(-2)&~6.546(-4)&-1.149(-3)&
   -4.573(-2)&-5.496(-3)&-7.864(-3)&-1.548(-3)&4.448(-3) \\
   2.50&0.1209~~~&-2.332(-2)&-7.605(-4)&-5.280(-3)&
   ~1.111(-4)&-5.723(-2)&-1.216(-2)&-4.555(-3)&1.313(-2) \\
   2.00&0.6365~~~&-2.618(-2)&-2.067(-2)&-1.992(-2)&
   ~7.509(-1)&-5.522(-1)&~2.515(-2)&-1.447(-2)&5.382(-2) \\
  \end{tabular}
}
 \end{center}
\end{table}%

\begin{table}
 \caption{Coefficients of the 1PN velocity potential and the Lagrangian
   displacement vectors associated with the binary motion shown along the
   equilibrium sequence of the irrotational Darwin-Riemann ellipsoid.
}
 \begin{center}
  \begin{tabular}{c|cccc|cc}
   $R/a_1$&$\t{e}$&$\t{f}$&$\t{g}$&$\t{h}$&
   $\t{S}_{21}$&$\t{S}_{22}$ \\ \hline
   6.00&0.1750&-2.520(-2)&-3.005(-2)&-2.997(-2)&
   1.465(-3)&-1.608(-3) \\
   5.50&0.2113&-2.517(-2)&-3.178(-2)&-3.164(-2)&
   2.264(-3)&-2.484(-3) \\
   5.00&0.2567&-2.465(-2)&-3.394(-2)&-3.368(-2)&
   3.649(-3)&-3.999(-3) \\
   4.50&0.3142&-2.320(-2)&-3.678(-2)&-3.627(-2)&
   6.185(-3)&-6.766(-3) \\
   4.00&0.3880&-1.991(-2)&-4.077(-2)&-3.970(-2)&
   1.116(-2)&-1.216(-2) \\
   3.50&0.4835&-1.278(-2)&-4.697(-2)&-4.450(-2)&
   2.175(-2)&-2.351(-2) \\
   3.00&0.6057&~3.204(-3)&-5.798(-2)&-5.167(-2)&
   4.663(-2)&-4.944(-2) \\
   2.50&0.7498&~4.035(-2)&-8.037(-2)&-6.307(-2)&
   0.1099~~~~&-0.1105~~~ \\
   2.00&0.8687&0.1120~~~&-0.1227~~~~&-8.151(-2)&
   0.2634~~~~&-0.2280~~~ \\
  \end{tabular}
 \end{center}
\end{table}%

\begin{table}
 \caption{The 1PN angular velocity, energy, and angular momentum in the
   ellipsoidal approximation along the equilibrium sequence of the
   irrotational Darwin-Riemann ellipsoid. The subscript ``${\rm e}$''
   denotes the case of the {\it ellipsoidal} approximation.
}
 \begin{center}
  \begin{tabular}{cc|ccc}
   $R/a_1$&$R_{\ast}/a_{\ast}$&$\d \t{\O}_{\rm e}$&$\t{E}_{\rm PNe}$&
   $\t{J}_{\rm PNe}$ \\ \hline
   6.00&6.071&0.1249&2.940(-2)&-0.7527 \\
   5.50&5.584&0.1401&5.961(-2)&-0.4835 \\
   5.00&5.103&0.1584&9.935(-2)&-0.1945 \\
   4.50&4.628&0.1808&0.1524~~~~&~0.1167 \\
   4.00&4.165&0.2084&0.2244~~~~&~0.4526 \\
   3.50&3.722&0.2426&0.3225~~~~&~0.8149 \\
   3.00&3.317&0.2837&0.4541~~~~&1.204~ \\
   2.50&2.992&0.3276&0.6165~~~~&1.626~ \\
   2.00&2.842&0.3527&0.7600~~~~&2.153~ \\
  \end{tabular}
 \end{center}
\end{table}%

\begin{table}
 \caption{Coefficients of the 1PN velocity potential in the ellipsoidal
   approximation along the equilibrium sequence of the irrotational
   Darwin-Riemann ellipsoid. The subscript ``${\rm e}$'' denotes the case of
   the {\it ellipsoidal} approximation.
}
 \begin{center}
  {\tabcolsep=3pt
  \begin{tabular}{c|cccc|cccc}
   $R/a_1$&$\t{p}_{\rm e}$&$\t{q}_{\rm e}$&$\t{r}_{\rm e}$&$\t{s}_{\rm e}$&
   $\t{e}_{\rm e}$&$\t{f}_{\rm e}$&$\t{g}_{\rm e}$&$\t{h}_{\rm e}$ \\ \hline
   6.00&-1.853(-2)&-6.155(-4)&~4.426(-5)&~4.445(-5)&
   0.1750&-2.537(-2)&-3.000(-2)&-2.997(-2) \\
   5.50&-2.157(-2)&-9.015(-4)&~5.892(-5)&~5.939(-5)&
   0.2113&-2.546(-2)&-3.168(-2)&-3.164(-2) \\
   5.00&-2.489(-2)&-1.365(-3)&~7.629(-5)&~7.751(-5)&
   0.2567&-2.519(-2)&-3.376(-2)&-3.368(-2) \\
   4.50&-2.791(-2)&-2.147(-3)&~8.957(-5)&~9.302(-5)&
   0.3141&-2.427(-2)&-3.642(-2)&-3.628(-2) \\
   4.00&-2.881(-2)&-3.529(-3)&~6.887(-5)&~7.960(-5)&
   0.3879&-2.220(-2)&-4.001(-2)&-3.972(-2) \\
   3.50&-2.218(-2)&-6.096(-3)&-1.043(-4)&-6.787(-5)&
   0.4833&-1.810(-2)&-4.518(-2)&-4.455(-2) \\
   3.00&~8.177(-3)&-1.107(-2)&-8.948(-4)&-7.645(-4)&
   0.6055&-1.052(-2)&-5.335(-2)&-5.185(-2) \\
   2.50&0.1068~~~&-2.057(-2)&-4.058(-3)&-3.635(-3)&
   0.7515&~2.764(-3)&-6.759(-2)&-6.380(-2) \\
   2.00&0.3336~~~&-3.347(-2)&-1.531(-2)&-1.414(-2)&
   0.8843&~2.684(-2)&-9.351(-2)&-8.377(-2) \\
  \end{tabular}
}
 \end{center}
\end{table}%

\section{Summary and discussion}

\subsection{Summary}

Using the scheme developed in a previous paper,\cite{TAS}
we have investigated the equilibrium sequences of irrotational and
incompressible binary systems in the 1PN approximation. Our results
presented in the previous section should be useful to check the numerical
code for solving the irrotational binary problem.
The conclusions are as follows.

\noindent
(1) Due to the 1PN effect, the orbital separation at the ISCO decreases
in proportion to the compactness parameter $M_{\ast}/c^2 a_{\ast}$.

\noindent
(2) The orbital angular velocity at the ISCO in units of
$\sqrt{M_{\ast}/a_{\ast}^3}$ has almost the same value as in 
the Newtonian case, even if we increase the compactness parameter.

\noindent
(3) The ellipsoidal approximation gives fairly accurate results
throughout the equilibrium sequence.

\noindent
(4) It is important to include the velocity potential at 1PN order.

\subsection{Discussion}

Finally, we discuss the velocity potential at 1PN order. As stated in a
previous paper,\cite{TAS} the internal velocity has an $x_3$ component.
Therefore, we suggest that when we solve an irrotational binary problem
numerically, we must include the $x_3$ component of the internal
velocity.

Next, we discuss the effect of the deformation at Newtonian order. Even
in the incompressible case, stars in the binary system deform from
ellipsoidal figures at Newtonian order.\cite{UE3} We can estimate the
contribution from the octupole moment on the angular velocity. First, we
assume that the deformation from an ellipsoidal figure is a small
perturbation. Accordingly, the deformation is expressed by the
Lagrangian displacement vectors (\ref{Lagdisp7}) and (\ref{Lagdisp8})
for the octupole moment. Here we neglect the contribution of the
hexadecapole and other moments as higher order terms. Then, the angular
velocity is written
\beqa
  \O^2 &=&{2M \over R^3} +{18\bI_{11} \over R^5} -{2 \over MR} \Bigl[ {2M
    \over R^3} +{18\bI_{11} \over R^5} +(2-F_a) F_a \O_{\rm DR}^2 \Bigr] 
  (\d I_1)_{\rm N} \nonumber \\
  &&-{4 \over R^6} \bigl[ 2(\d I_{111})_{\rm N} -3(\d
  I_{122})_{\rm N} -3(\d I_{133})_{\rm N} \bigr] \nonumber \\
  &&+{1 \over c^2} \d \O^2 +O(R^{-7}), \label{higherangvelo}
\eeqa
where the subscript N denotes the perturbation of Newtonian
order. Equation (\ref{higherangvelo}) is approximated as
\beqa
  \O^2 &=&{2 M \over R^3} \Bigl[ 1+{3 \over 5 \t{R}^2} (2 -\a_2^2
  -\a_3^2) -\omega_1 \t{t}_{21} -\omega_2 \t{t}_{22} +C_{\rm s} {\t{R}^3 
    \over 2\a_2 \a_3} \d \t{\O}^2 \Bigr],
\eeqa
where $\t{t}_{21}$ and $\t{t}_{22}$ denote the amplitude of the
Lagrangian displacement vectors (\ref{Lagdisp7}) and (\ref{Lagdisp8}),
and
\beqa
  \omega_1 &=& {1 \over 5\t{R}} \Bigl[ 1+{3 \over 5 \t{R}^2} (2 -\a_2^2
  -\a_3^2) \Bigr] \bigl[ 1+ (2-F_a) F_a \bigr] \nonumber \\
  &&+{6 \over 35\t{R}^3} \Bigl( 3 -{1 \over 2} \a_2^2 +{3 \over 2}
  \a_3^2 \Bigr), \\
  \omega_2 &=& -{12 \over 35 \t{R}^3} (\a_2^2 -\a_3^2).
\eeqa
{}From Table II, we can obtain the values in the range of $0.01 \le
C_{\rm s} \le 0.05$ as
\beqa
  &&\omega_1 \t{t}_{21} \sim 0.14 \t{t}_{21}, \\
  &&\omega_2 \t{t}_{22} \sim 0.001 \t{t}_{22}, \\
  &&0.0243 \le C_{\rm s} {\t{R}^3 \over 2\a_2 \a_3} \d \t{\O}^2 \le
  0.121.
\eeqa
Therefore, if the amplitude $\t{t}_{21}$ (and $\t{t}_{22}$) is much less
than 1, i.e., in the range of $0 \le \t{t}_{21} \le 0.1$, the
contribution of the octupole deformation to the angular velocity is
smaller than that of the 1PN correction. On the other hand, if
$\t{t}_{21}$ and/or $\t{t}_{22}$ have values near or greater than 1, the
above perturbative treatments break down. In this case, we cannot
estimate the contribution of the octupole moment until we construct the
equilibrium figure including the octupole moment.

Moreover, we would like to mention the importance of the velocity
potential of 1PN order. If we neglect it, we obtain different features
for the ISCO. Accordingly, we must take into account the velocity
potential at 1PN order even when we use the ellipsoidal approximation.

{}From Fig. 5, we can estimate the frequency of gravitational waves
at the ISCO as
\beqa
  450~ [{\rm Hz}] \lsim
  \Bigl( {M_{\ast} \over 1.6M_{\odot}} \Bigr) f_{\rm GW}
  \lsim 800~ [{\rm Hz}],
\eeqa
if we consider that we have $0 \lsim n \lsim 1$ for the
polytropic indicies and $0.15 \lsim C_{\rm s} \lsim 0.25$ for the
compactness parameter of neutron stars.

\section*{Acknowledgements}

I would like to thank H. Asada, Y. Eriguchi, T. Nakamura, K. Nakao and
M. Shibata for useful discussions, and I would also like to thank H.
Sato and N. Sugiyama for helpful comments and continuous encouragement.
This work was partly supported by a Grant-in-Aid for Scientific Research
Fellowship (No.9402) of the Japanese Ministry of Education, Science,
Sports and Culture.

\appendix
\section{Explicit Form of $\d U$} \label{deltaUform}

The correction of the self-gravity potential induced by the deformation
of the binary system is given by\cite{chandra71,TS}
\beqa
  \d U ={1 \over c^2} \sum_{ij} S_{ij} \d U^{(ij)},
\eeqa
where
\beqa
  \d U^{(ij)} =-\sum_k {\p \over \p x_k} \int d^3 x' {\r(x')
    \xi_k^{(ij)} (x') \over |x-x'|}. \label{deltaU}
\eeqa
Substituting the Lagrangian displacement vectors into
Eq. (\ref{deltaU}), we separately obtain $\d U$ as
\beqa
  \d U=\d U^{1 \ra 1} +\d U^{2 \ra 1},
\eeqa
where
\beqa
  \d U^{1 \ra 1} &=&{1 \over c^2} \Bigl[ S_{11} (D_{3,3} -D_{1,1})
  +S_{12} (D_{3,3} -D_{2,2}) +S_{31} \Bigl( D_{112,2} -{1 \over 3}
  D_{111,1} \Bigr) \nonumber \\
  & &\hspace{20pt}+S_{32} \Bigl( D_{223,3} -{1 \over 3} D_{222,2}
  \Bigr) +S_{33} \Bigl( D_{133,1} -{1 \over 3} D_{333,3} \Bigr)
  \nonumber \\
  & &\hspace{20pt}-{1 \over 2} S_0 U^{1 \ra 1}_{,1} +S_{21} \Bigl
  ( D_{13,3} -{1 \over 2} D_{11,1} \Bigr) +S_{22} (D_{13,3} -D_{12,2})
  \Bigr], \\
  \d U^{2 \ra 1} &=&{1 \over c^2} \Bigl[ S_{11} (D_{3,3}^{2 \ra 1}
  -D_{1,1}^{2 \ra 1}) +S_{12} (D_{3,3}^{2 \ra 1} -D_{2,2}^{2 \ra 1})
  +S_{31} \Bigl( D_{112,2}^{2 \ra 1} -{1 \over 3} D_{111,1}^{2 \ra 1}
  \Bigr) \nonumber \\
  & &\hspace{20pt}+S_{32} \Bigl( D_{223,3}^{2 \ra 1} -{1 \over 3}
  D_{222,2}^{2 \ra 1} \Bigr) +S_{33} \Bigl( D_{133,1}^{2 \ra 1} -{1
  \over 3} D_{333,3}^{2 \ra 1} \Bigr) \nonumber \\
  & &\hspace{20pt}+{1 \over 2} S_0 U^{2 \ra 1}_{,1} -S_{21}
  \Bigl( D_{13,3}^{2 \ra 1} -{1 \over 2} D_{11,1}^{2 \ra 1} \Bigr)
  -S_{22} (D_{13,3}^{2 \ra 1} -D_{12,2}^{2 \ra 1}) \Bigr].
\eeqa
Note that the sign of $\xi_i$ for $\d U^{2 \ra 1}$ is opposite to that
of $\d U^{1 \ra 1}$ if the components of $\xi_i$ are even functions
of $x_i$.

\section{Perturbed Terms} \label{perturbedterm}

The definitions of $\d I_i$, $\d I_{ij}$, and $\d W$ are
\beqa
  \d I_i &\equiv& \int d^3 x \r \xi_i, \\
  \d I_{ij} &\equiv& \int d^3 x \r \sum_l \xi_l {\p \over \p x_l} (x_i
  x_j), \\
  \d W &\equiv& \sum_{i=1}^3 \d W_{ii},
\eeqa
where
\beqa
  \d W_{ij} = \pi \r \bigl( -2B_{ij} \d I_{ij} +a_i^2 \d_{ij}
  \sum_l A_{il} \d I_{ll} \bigr).
\eeqa
The components are
\beqa
  \d I_1 &=&{1 \over 2c^2} (M S_0 +S_{21} I_{11}), \\
  \d I_{11} &=& {2 \over c^2} \Bigl( S_{11} I_{11} +{1 \over 3} S_{31}
  I_{1111} -S_{33} I_{1133} \Bigr), \\
  \d I_{22} &=& {2 \over c^2} \Bigl( S_{12} I_{22} -S_{31} I_{1122} +{1
    \over 3} S_{32} I_{2222} \Bigr), \\
  \d I_{33} &=& {2 \over c^2} \Bigl( -S_{11} I_{33} -S_{12} I_{33}
  -S_{32} I_{2233} +{1 \over 3} S_{33} I_{3333} \Bigr), \\
  \d W_{11} &=& {2\pi \r \over c^2} \biggl[ S_{11} \Bigl\{ (a_1^2 A_{11}
  -2B_{11}) I_{11} -a_1^2 A_{13} I_{33} \Bigr\} \nonumber \\
 & &\hspace{25pt} +S_{12} (a_1^2 A_{12}
  I_{22} -a_1^2 A_{13} I_{33}) \nonumber \\
  & &\hspace{25pt}+S_{31} \Bigl\{ {1 \over 3} (a_1^2 A_{11}
  -2B_{11}) I_{1111} -a_1^2 A_{12} I_{1122} \Bigr\} \nonumber \\
  & &\hspace{25pt}+S_{32} \Bigl( {1
    \over 3} a_1^2 A_{12} I_{2222} -a_1^2 A_{13} I_{2233} \Bigr)
  \nonumber \\
  & &\hspace{25pt}+S_{33} 
  \Bigl\{ -(a_1^2 A_{11} -2B_{11}) I_{1133} +{1 \over 3} a_1^2 A_{13}
  I_{3333} \Bigr\} \biggr], \\
  \d W_{22} &=& {2\pi \r \over c^2} \biggl[ S_{11} (a_2^2 A_{12} I_{11}
  -a_2^2 A_{23} I_{33}) \nonumber \\
  & &\hspace{25pt} +S_{12} \Bigl\{ (a_2^2 A_{22} -2B_{22}) I_{22}
  -a_2^2 A_{23} I_{33} \Bigr\} \nonumber \\
  & &\hspace{25pt} +S_{31} \Bigl\{ {1 \over 3} a_2^2 A_{12}
  I_{1111} -(a_2^2 A_{22} -2B_{22}) I_{1122} \Bigr\} \nonumber \\
  & &\hspace{25pt} +S_{32} \Bigl\{ {1
    \over 3} (a_2^2 A_{22} -2B_{22}) I_{2222} -a_2^2 A_{23} I_{2233}
  \Bigr\} \nonumber \\
  & &\hspace{25pt}+S_{33} \Bigl( -a_2^2 A_{12} I_{1133} +{1 \over 3} a_2^2
  A_{23} I_{3333} \Bigr) \biggr], \\
  \d W_{33} &=& {2\pi \r \over c^2} \biggl[ S_{11} \Bigl\{ a_3^2 A_{13}
  I_{11} -(a_3^2 A_{33} -2B_{33}) I_{33} \Bigr\} \nonumber \\
  & &\hspace{25pt} +S_{12} \Bigl\{ a_3^2
  A_{23} I_{22} -(a_3^2 A_{33} -2B_{33}) I_{33} \Bigr\} \nonumber \\
  & &\hspace{25pt}+S_{31} \Bigl( {1 \over 
    3} a_3^2 A_{13} I_{1111} -a_3^2 A_{23} I_{1122} \Bigr) \nonumber \\
  & &\hspace{25pt} +S_{32}
  \Bigl\{ {1 \over 3} a_3^2 A_{23} I_{2222} -(a_3^2 A_{33} -2B_{33})
  I_{2233} \Bigr\} \nonumber \\
  & &\hspace{25pt}+S_{33} \Bigl\{ -a_3^2 A_{13} I_{1133} +{1 \over 3}
  (a_3^2 A_{33} -2B_{33}) I_{3333} \Bigr\} \biggr].
\eeqa



\begin{thebibliography}{99}
\bibitem{LIGO}
A. Abramovici et al., Science {\bf 256} (1992), 325.

\bibitem{VIRGO}
C. Bradaschia et al., Nucl. Instrum. and Methods {\bf A289} (1990), 518.

\bibitem{GEO}
J. Hough, in {\it Proceedings of the Sixth Marcel Grossmann Meeting},
ed. H. Sato and T. Nakamura
(World Scientific, Singapore, 1992), p. 192.

\bibitem{TAMA}
K. Kuroda et al., in {\it Proceedings of International Conference on
Gravitational Waves: Sources and Detectors}, ed. I. Ciufolini and
F. Fidecaro (World Scientific, Singapore, 1997), p. 100.

\bibitem{Thorne} For example, K. S. Thorne, in {\it 
Proceeding of Snowmass 94 Summer Study on Particle and Nuclear 
Astrophysics and Cosmology}, ed. E. W. Kolb and R. Peccei 
(World Scientific, Singapore, 1995) and references cited therein. 

\bibitem{lindblom}
L. Lindblom, Astrophys. J. {\bf 398} (1992), 569.

\bibitem{shibata}
M. Shibata, Phys. Rev. {\bf D55} (1997), 6019;
Prog. Theor. Phys. {\bf 96} (1996), 317. \\ 
M. Shibata, K. Taniguchi, and T. Nakamura,
Prog. Theor. Phys. Supple. No.~128 (1997), 295.

\bibitem{BCSST} 
T. W. Baumgarte, G. B. Cook, M. A. Scheel, S. L. Shapiro
and S. A. Teukolsky,
Phys. Rev. Lett. {\bf 79} (1997), 1182; Phys. Rev. {\bf D57} (1998), 7299
and references cited therein. 

\bibitem{Kochanek}
C. S. Kochanek, Astrophys. J. {\bf 398} (1992), 234.

\bibitem{BC}
L. Bildsten and C. Cutler, Astrophys. J. {\bf 400} (1992), 175.

\bibitem{UE1}
K. Ury\=u and Y. Eriguchi, Mon. Not. R. Astron. Soc. {\bf 282} (1996),
653.

\bibitem{UE2}
K. Ury\=u and Y. Eriguchi, Mon. Not. R. Astron. Soc. {\bf 296} (1998),
L1; Astrophys. J. Supple. Ser. {\bf 118} (1998), 563.

\bibitem{UE3}
K. Ury\=u and Y. Eriguchi, Mon. Not. R. Astron. Soc. {\bf 299} (1998),
575.

\bibitem{TNLWOK}
K. Taniguchi and T. Nakamura, Prog. Theor. Phys. {\bf 96} (1996),
693. \\
D. Lai and A. G. Wiseman, Phys. Rev. {\bf D54} (1996), 3958. \\
W. Ogawaguchi and Y. Kojima, Prog. Theor. Phys. {\bf 96} (1996), 901.

\bibitem{WMM}
J. R. Wilson and G. J. Mathews, Phys. Rev. Lett. {\bf 75} (1995), 4161. \\
J. R. Wilson, G. J. Mathews and P. Marronetti, Phys. Rev. {\bf D54}
(1996), 1317. \\
G. J. Mathews, P. Marronetti and J. R. Wilson, Phys. Rev. {\bf D58}
(1998), 043003. \\
G. J. Mathews and J. R. Wilson, Astrophys. J. {\bf 482} (1997), 929.

\bibitem{density}
D. Lai, Phys. Rev. Lett. {\bf 76} (1996), 4878. \\
A. G. Wiseman, Phys. Rev. Lett. {\bf 79} (1997), 1189. \\
P. R. Brady and S. A. Hughes, Phys. Rev. Lett. {\bf 79} (1997), 1186. \\
\'E. \'E. Flanagan, Phys. Rev. {\bf D58} (1998), 124030. \\
K. S. Thorne, Phys. Rev. {\bf D58} (1998), 124031.

\bibitem{Lom}
J. C. Lombardi, F. A. Rasio and S. L. Shaipro, 
Phys. Rev. {\bf D56} (1997), 3416. 

\bibitem{TAS}
K. Taniguchi, H. Asada and M. Shibata, Prog. Theor. Phys. {\bf 100}
(1998), 703.

\bibitem{TS}
K. Taniguchi and M. Shibata, Phys. Rev. {\bf D56} (1997), 798. \\
M. Shibata and K. Taniguchi, ibid. {\bf 56} (1997), 811. 

\bibitem{BGM}
S. Bonazzola, E. Gourgoulhon and J.-A. Marck, Phys. Rev. {\bf D56} (1997),
7740.

\bibitem{Asada}
H. Asada, Phys. Rev. {\bf D57} (1998), 7292.

\bibitem{Shibata98}
M. Shibata, Phys. Rev. {\bf D58} (1998), 024012. 

\bibitem{Teukolsky}
S. A. Teukolsky, Astrophys. J. {\bf 504} (1998), 442.

\bibitem{Gourgoulhon}
E. Gourgoulhon, preprint gr-qc/9804054.

\bibitem{chandra65}
S. Chandrasekhar, Astrophys. J. {\bf 142} (1965), 1488.

\bibitem{chandra67}
S. Chandrasekhar, Astrophys. J. {\bf 142} (1965), 1513; {\bf 147}
(1967), 334; {\bf 167} (1971), 447.

\bibitem{chandra71}
S. Chandrasekhar, Astrophys. J. {\bf 148} (1967), 621; {\bf 167}
(1971), 455.

\bibitem{chandra74}
S. Chandrasekhar and D. D. Elbert, Astrophys. J. {\bf 192} (1974), 731;
{\bf 220} (1978), 303.

\bibitem{chandra69}
S. Chandrasekhar, {\it Ellipsoidal Figures of 
Equilibrium} (Yale University Press, New Haven, CT, 1969). 

\bibitem{Aizenman}
M. L. Aizenman, Astrophys. J. {\bf 153} (1968), 511.

\bibitem{LRS94}
D. Lai, F. A. Rasio and S. L. Shapiro, Astrophys. J. {\bf 420} (1994),
811.

\bibitem{LRS93}
D. Lai, F. A. Rasio and S. L. Shapiro, Astrophys. J. Supple. Ser. {\bf
  88} (1993), 205.

\bibitem{Bardeen}
J. M. Bardeen, Astrophys. J. {\bf 167} (1971), 425.

\bibitem{SZ}
S. L. Shapiro and S. Zane, Astrophys. J. Supple. Ser. {\bf 117} (1998),
531.

\end{thebibliography}
\end{document}